\documentclass[10pt,twocolumn,numbers]{article}

\usepackage{graphicx}
\usepackage{amssymb}
\usepackage{amsthm,amsmath,dsfont}
\usepackage{mathtools}

\usepackage{lineno}

\usepackage[colorlinks=true, citecolor=blue]{hyperref}
\usepackage[linesnumbered,lined,ruled,commentsnumbered]{algorithm2e}
\usepackage{algcompatible}

\usepackage{wrapfig}
\usepackage[numbers,sort&compress]{natbib}
\usepackage{comment}
\usepackage{longtable}
\usepackage{multirow}
\usepackage{bbding}
  
\usepackage[bottom]{footmisc}
\usepackage{booktabs}
\usepackage{pifont}
\usepackage{float}
\usepackage{subfigure}
\usepackage{wrapfig}
\usepackage{framed,xcolor}
\usepackage{newfloat}
\usepackage[xcolor,framemethod=TikZ]{mdframed}
\usepackage{amsthm}
\usepackage{thmtools}
\usepackage{authblk}
\patchcmd{\thebibliography}{\section*{\refname}}{}{}{}
\usepackage{tikz}
\usetikzlibrary{arrows,shapes,chains,calc,patterns,decorations.pathreplacing}
\usepackage{stmaryrd}
\usepackage{fancyhdr}
\usepackage{subfigure}
\usepackage{comment}
\def\BibTeX{{\rm B\kern-.05em{\sc i\kern-.025em b}\kern-.08em
		T\kern-.1667em\lower.7ex\hbox{E}\kern-.125emX}}

\renewcommand{\headrulewidth}{2pt}

\newlength\FHoffset

\fancyheadoffset{\FHoffset}

\newlength\FHleft
\newlength\FHright

\newbox\FHline
\setbox\FHline=\hbox{\hsize=\paperwidth%
	\hspace*{\FHleft}%
	\rule{\dimexpr\headwidth-\FHleft-\FHright\relax}{\headrulewidth}\hspace*{\FHright}%
}

\newcommand{\RR}{\mathbb{R}}

\newcommand{\dd}{\mathrm{d}}

\newcommand{\A}{\mathcal{A}}
\newcommand{\B}{\mathcal{B}}
\newcommand{\C}{\mathcal{C}}

\newtheoremstyle{theoremdd}
{\topsep}
{\topsep}
{\itshape}
{0pt}
{\fontfamily{cmss}\selectfont\bfseries}
{.}
{ }
{\thmname{#1}\thmnumber{ #2}\thmnote{ (#3)}}

\theoremstyle{theoremdd}

\mdfdefinestyle{myStyle}{roundcorner=5pt,backgroundcolor=brown!20,linecolor=red!40!black,linewidth=1.5pt}

\usepackage{titlesec}
\titleformat*{\section}{\fontfamily{cmss}\selectfont\large\bfseries\color{red!40!black}}
\titleformat*{\subsection}{\fontfamily{cmss}\selectfont\normalsize\bfseries\color{red!40!black}}
\titleformat*{\subsubsection}{\fontfamily{cmss}\selectfont\normalsize\color{red!40!black}}
\usepackage[left=0.69 in, right=0.69 in, top=1.1 in, bottom=1.3 in]{geometry}
\graphicspath{{./Figures/}}

\newcommand\blfootnote[1]{%
	\begingroup
	\renewcommand\thefootnote{}\footnote{#1}%
	\addtocounter{footnote}{-1}%
	\endgroup
}
\renewcommand\abstractname{\fontfamily{cmss}\selectfont\normalsize\bfseries\color{red!40!black}\textbf{Abstract}}

\renewenvironment{abstract}{%
	\centering\small
	\list{}{\leftmargin1.5cm \rightmargin\leftmargin}
	\item\relax
	
	\begin{mdframed}[]
		\item[\hskip\labelsep\scshape\abstractname.]%
	}{%
	\end{mdframed}
	\endlist \par\bigskip
}
\makeatletter
\patchcmd{\@maketitle}{\LARGE \@title}{\fontfamily{cmss}\selectfont\LARGE\color{red!40!black}\@title}{}{}
\makeatother

\graphicspath{{./Figures/}}

\begin{document}
	
		
		
		\title{\fontfamily{cmss}\selectfont Pay for intersection priority: A free market mechanism for connected vehicles}
		


\author[1]{DianChao Lin}
\author[1,2]{Saif Eddin Jabari$^\dagger$}

\affil[1]{New York University Tandon School of Engineering, Brooklyn NY}
\affil[2]{New York University Abu Dhabi, Saadiyat Island, P.O. Box 129188, Abu Dhabi, U.A.E.}

\date{}

\twocolumn[
\begin{@twocolumnfalse}
	
\maketitle	

{ \fontfamily{cmss}\selectfont\large\bfseries		
\begin{abstract}
{ \normalfont\normalsize
	The rapid development and deployment of vehicle technologies offer opportunities to re-think the way traffic is managed.  This paper capitalizes on vehicle connectivity and proposes an economic instrument and corresponding cooperative framework for allocating priority at intersections.  The framework is compatible with a variety of existing intersection control approaches. Similar to free markets, our framework allows vehicles to trade their time based on their (disclosed) value of time. We design the framework based on transferable utility games, where winners (time buyers) pay losers (time sellers) in each game.  We conduct simulation experiments of both isolated intersections and an arterial setting. The results show that the proposed approach benefits the majority of users when compared to other mechanisms both ones that employ an economic instrument and ones that do not.  We also show that it drives travelers to estimate their value of time correctly, and it naturally dissuades travelers from attempting to cheat.
	
	\bigskip
	
	\textbf{\fontfamily{cmss}\selectfont Keywords}: Intersection operations, economic instruments, cooperative games, connected vehicles, value of time (VOT), direct-transaction, side payment.
	
	\bigskip
}
\end{abstract}
}
\end{@twocolumnfalse}
]
		
		
	
	
	
\section{Introduction}
Intersections\blfootnote{$\dagger$ Corresponding author, E-mail: \url{sej7@nyu.edu}} are the bottlenecks of a city's traffic network. The interaction between multiple conflicting streams of traffic at intersections presents safety, efficiency, and environmental challenges. To address these challenges, intersection controllers schedule the priorities of different traffic streams in a way that optimizes one or more operational metrics. In this context, compared to a traditional vehicle (TV) environment, a system consisting of connected vehicles (CVs) can collect and utilize more detailed data and exchange information between vehicles (and the Internet) efficiently.  This creates opportunities to rethink the way intersections are controlled. 

In practice, priority is traditionally provided to competing vehicles, modes, or infrastructure elements, with the purpose of improving the performance of the system.  For example, capacity conversion factors are used to differentiate vehicles belonging to different classes  \cite{manual2010hcm2010} and major streets usually have higher priority than minor streets. To promote transit, one prioritizes vehicles with larger numbers of passengers and distinguished vehicles by trip purpose, e.g., high-occupancy vehicle lanes and bus-only lanes \cite{hu2015coordinated, furth2000conditional, eichler2006bus, ma2013dynamic, ma2014integrated}.  Considering different travel time requirements, toll lanes are designed to allow paying vehicles to go faster.  But such toll lanes are usually expensive \cite{dahlgren2002high} and they typically do not apply tiered pricing for different travelers because of limited lane resources.

In a CV environment, controllers can easily account for some heterogeneous characteristics of the vehicles, e.g., transit signal priority \cite{hu2015coordinated}. Furthermore, CV technology allows vehicles to offer detailed trip information to the traffic system in advance, including trip purposes and trip urgency levels. 
Control tools, in turn, can use such rich information to schedule vehicles through network intersections in ways that best benefit the users.  Researchers often use value of time (VOT) to quantify ``urgency'', and this is typically done with the purpose of assigning intersection priorities (by an operator via controllers in the infrastructure) to optimize \emph{valuated} travel times.  But VOT can only be known to (or estimated by) the vehicles themselves (more accurately, their occupants). There is no way for infrastructure controllers to know whether the VOTs that are being reported are true or not. This can lead to higher priorities being given to vehicles with overestimated (or deliberately inflated) VOTs.  This incentivizes more false reporting by users that wish to take advantage of the system and it encourages abandonment on part of those users that are honest in their VOT reporting. Such a  system will degrade to a traditional one rapidly due to the asymmetric information about VOT, which is similar to Akerlof's market for ``lemons" \cite{akerlof1978market}. 

The combination of economic instruments with traffic control are quickly becoming a popular way to address this issue \cite{schepperle2007traffic,schepperle2007agent,schepperle2008auction,vasirani2012market, carlino2013auction, levin2015intersection,lloret2016envy,lin2019transferable}.  They are known to be powerful traffic management tools, both theoretically and in practice (e.g., tolls, public transport subsidies, and vehicle license auctions).  CVs will reinforce this trend and allow for new ways to incorporate economic instruments in traffic management.  Combined with \emph{quick mobile payment} technologies using \emph{digital wallets} (or e-wallets), economic instruments can be used to manage demand effectively, guide trips, and improve social welfare. For intersection management, economic instruments with appropriately designed rules for transactions can cater to the heterogeneous needs of the different users (e.g., based on VOT) while also increasing social benefit. 

To this end, research has emerged on auction-based intersection operations over the past decade. When passing an ``auction-required'' intersection, vehicles need to auction their passing rights with others from conflicting traffic streams (vehicles about to leave) based on their valuated delay or travel time \cite{schepperle2007agent,schepperle2008auction,vasirani2012market, carlino2013auction, levin2015intersection,li2020user}.  However, auctions (both first and second price auctions) are not revenue neutral (budget balanced) instruments.  They involve an operator (e.g., governmental or private entity) that charges vehicles for the service, which is not a \emph{pareto-improving} mechanism as it puts the user in a disadvantaged position relative to the operator (operators in this case play a role that is similar to \emph{sink agents} described in  \cite{nath2019efficiency}).  Furthermore, because VOT is correlated with individual welfare, such mechanisms can enlarge the rich-poor gap. In order to solve these problems, some researchers proposed credit auction schemes \cite{carlino2013auction}, which are also used in traffic network management \cite{yang2011managing}. The credit is like a new currency, which is pre-distributed by the operator freely and periodically. Citizens use the credits for traffic auctions.  Hence, it is revenue-neutral but in an indirect way. However, this mechanism is time-consuming, and it may render travel a more complicated process \cite{lin2020comparative}, which makes it impractical. Some researchers proposes a mechanism (referred to as \emph{direct-transactions} in this paper) similar to a ``free-market'' at the intersection. Vehicles with heterogeneous VOTs can engage in trading priority for direct monetary compensation, i.e., they can sell or buy priority to/from others   \cite{lloret2016envy, lin2019transferable}. Although these works demonstrated that direct transactions are Pareto-improving and practical, they failed to reveal a critical advantage: their compatibility with existing control technique.

To address these drawbacks, this paper proposes a simple cooperative framework for allocating priority at intersections. The proposed economic framework can be combined with multiple control approaches and (thus) various types of signalized intersections.  Unlike the optimization framework proposed in \cite{lloret2016envy}, we employ a cooperative games approach, namely \emph{transferable utilities} (TU) games. Our approach involves an intersection controller capable of communicating with participating vehicles its \emph{control zone}. The controller develops a discharge strategy based on the VOTs of the vehicles, which the vehicles communicate to the controller, and a control objective, which can differ by controller type.  Based on the discharge strategy, vehicles are split into three groups: \emph{payers}, \emph{payees}, and \emph{indifferent vehicles} that play a TU game. The TU game determines the total payment that is to be transferred from the payers to the payees. The mechanism is quite simple. Vehicles only need to determine their VOTs and route information before travel. This also facilitates quick and simple transactions, which is needed in the context of real-time intersection operations.  The proposed framework is compatible with existing control techniques and can, therefore, be viewed as a complementary framework, one that is meant to improve upon aspects of existing control frameworks, not compete with them.

This paper extends our previous work on transferable utility games for intersection control \cite{lin2019transferable} in the following ways: (i)  we outline  three popular control methods that are compatible with our game framework, (ii) we apply our game framework to more complex intersection layouts in our experiments, (iii) we explore the impact of travelers misreporting their VOTs in one of our experiments, and (iv) we include an arterial implementation in our experiments section. The next section reviews relevant literature, namely VOT and intersection control, mainly for a CV environment. The following section outlines the proposed approach and how it can be incorporated with some existing control methods. This is followed by simulation experiments. The last section concludes the paper.

\section{Literature Review}
\label{S:LReview}

\subsection{Value of Time (VOT)}
\label{SS:TravelDemand}
VOT serves as measure of heterogeneity among the vehicles at an intersection, it has been regarded a means to compensate for variation \cite{varian1992microeconomic}. It captures what different travelers are willing to pay for for travel time savings and has been investigated for more than half a century \cite{west1946economic, moses1963value, beesley1965value, becker1965theory}. In the past 20 years, studies of VOT were specialized to different parts of the world such as Japan \cite{kato2010meta}, America \cite{brownstone2005valuing} and Europe \cite{shires2009international}.

VOTs are represented by econometric models that include personal information about the travelers \cite{small2007economics}. It, therefore, requires input from the travelers themselves and cannot be estimated by roadside controllers. It can, however, be easily estimated by a computer onboard the travelers CV with input from the traveler.  We refer to \cite{lin2020comparative} for more details on how such a process may be developed.

\subsection{Economic Instruments for Intersection Control}
Intersection control tools in the literature that use economic instruments generally aim to optimize valuated travel time or valuated delay.  Currently, there are three transaction mechanisms in the literature: auctions, credit auctions, and direct-transactions.

Auctions are the most popular tools for deciding on how payments are made in these systems. Using a second-price sealed-bid auction, Schepperle and B{\"o}hm \cite{schepperle2007agent,schepperle2008auction} optimized the discharge order of vehicles.  Their mechanism was shown to outperform a first-come-first-served discharge policy in terms of mean valuated travel time.  Vasirani and Ossowski \cite{vasirani2009market,vasirani2011computational,vasirani2012market} proposed a reservation-based controller and a Walrasian auction \cite{walras2014leon} to simulate intersection traffic. They found that when the volume is high, optimizing the valuated time saved will significantly increase the average delay. Levin and Boyles \cite{levin2015intersection} found a similar shortcoming  using a first-price auction. To address this, Gende \cite{gende2015using} proposed an unbalanced control method, where high volume approaches use an auction-based scheme (first-price), and other approaches use a traditional controller. This significantly reduced the average delay. It is worth noting that there exist intersection control techniques that employ \emph{consensus-based} algorithms that resemble auctions (e.g., \cite{choi2009consensus,molinari2018automation,raphael2017intersection}). But these do not involve monetary transactions, and while they are interesting, they are not reviewed here.

If the operator uses credits (or tokens) to replace real currency, revenue-neutrality can be realized.  Carlino et al. \cite{carlino2013auction} extended the reservation-based control to a second-price credit auction for autonomous vehicles (AVs).  They allowed vehicles to abandon the system (with 0 bid) and following vehicles in the system would take their place.  Mashayekhi and List \cite{mashayekhi2015multiagent} let movements (groups of non-conflicting approaches at the intersection) bid for their priorities. They employed reinforcement learning techniques, and a first-price auction was chosen for reward calculation. 

Direct transactions (priced exchanges) were proposed with the goal of achieving revenue neutrality \cite{lloret2016envy, lin2019transferable}.  Schepperle et al. \cite{schepperle2007traffic,schepperle2007towards} first proposed a similar idea for trading priorities at intersections (via time-slot exchange and compensation). Their compensation mechanism is simple: payments are equal to VOT (currency units per second of reduced waiting time).  Lloret-Batlle and Jayakrishnan \cite{lloret2016envy} tried to pacify user envy when deciding on the payments.  Here, \emph{envy} is when a user prefers another user's allocation under the given price more than their own allocation.  \emph{Envy-freeness} is a sort of equilibrium allocation that is difficult to realize in practice, the authors of \cite{lloret2016envy} aimed to minimize it instead. They applied the idea to a dual-ring phase control with a rolling horizon.

\subsection{Intersection Control for CV without Economic Instruments}
\label{SS:IntersectionControl}
For the sake of completeness, we also review traditional (i.e., without economic instruments) intersection control approaches developed for CVs.

Researchers have proposed various optimization objectives to improve intersection performance for CVs. Delay minimization is the most widely applied objective in intersection control \cite{hu2015coordinated, guler2014using, yang2016isolated, yang2017multi, goodall2013traffic, feng2015real, jin2012multi, li2014signal, adkinsgame, zohdy2012game, bui2017game, bui2018cooperative, stone2015autonomous, dresner2005multiagent, dresner2004multiagent}, and it can be simply calculated as the difference between real-speed travel time and free-flow-speed travel time.  Moreover, minimizing personal delay can be seen as an implicit objective for reservation-based systems, such as first-in-first-out systems \cite{stone2015autonomous, dresner2004multiagent}.  Other papers minimized travel time or maximized travel speed, which are (in essence) equivalent to minimizing delay \cite{jin2012advanced, zhu2015linear, lee2013cumulative, sharon2017network}. 

Other recent approaches focus on minimizing queue sizes \cite{feng2015real, alvarez2010game, castillo2015solving}.  Both traditional actuated control techniques and recent decentralized control approaches can be seen as falling in this category as well \cite{wongpiromsarn2012distributed,varaiya2013max,gregoire2014back,xiao2014pressure,gregoire2015capacity,gregoire2016back,li2019position,rey2019blue,li2020backpressure}.  Fuel consumption is also a common optimization index \cite{rakha2011eco, li2015eco, zhang2016optimal, kamalanathsharma2016leveraging}, which combines environmental considerations with economic efficiency and resource utilization. When considering both efficiency and comfort, number-of-stops or acceleration/deceleration are useful indices \cite{guler2014using, yang2016isolated, gueriau2016assess, elhenawy2015intersection}. Other approaches have focused on safety, e.g., Lu et al. \cite{lu2014rule} set collision-avoidance as the goal and performed field tests using a rule-based control algorithm. Other researchers applied multi-objective techniques for intersection optimization 
\cite{goodall2013traffic, khondaker2015variable}.

\section{Mechanism Design} %
\begin{figure}[ht!]
	\scriptsize
	\tikzstyle{format}=[rectangle,draw,thin,fill=white,align=center]
	\begin{center}
		\begin{tikzpicture}[node distance=1cm,
			auto,>=latex',
			thin,
			start chain=going below,
			every join/.style={norm}]
			\node[format](n0){\small Determine $\mathcal{S}^*(t)$ and $T^*(t)$};
			\node [draw, diamond, aspect=2,below of=n0,text width=2cm,align=center,yshift=-.7cm] (n1){\small Is $\mathcal{S}^*(t)$ better than $\mathcal{S}(t)$?};
			\node[format,below of=n1,text width=3.5cm,yshift=-0.9cm] (n2){\footnotesize Determine groups $\A$ (payers) and $\B$ (payees)};
			\node[format,below of=n2, text width=3cm,yshift=-.2cm] (n3){\footnotesize $\A$ and $\B$ play a TU game to determine $\sigma$};
			\node[format,below of=n3,text width=3.8cm,yshift=-.2cm] (n4){\footnotesize Calculate $\sigma$ and split $\sigma$/$-\sigma$ among payers/payees};
			\node[format,below of=n4,text width=3cm,yshift=-.2cm](n7){\footnotesize Wait until next update time};
			\draw[->] (n0.south) -- (n1);
			\draw[->] (n1.south) -- node[anchor=center,xshift=-.3cm] {\footnotesize Yes} (n2);
			\draw[->] (n2.south) -- (n3);
			\draw[->] (n1.east) -- node[anchor=center,above] {\footnotesize No}(2.5,-1.7) -- (2.5,-7.2) -- (n7);
			\draw[->] (n3.south) -- (n4);
			\draw[->] (n4.south) -- (n7);
			\draw[->] (n7.west) -- (-2.5,-7.2) -- (-2.5,0.0) -- (n0);
			\node[format,right of=n0,draw = white,text width=1.75cm,xshift = 3.15cm,](n0r){\footnotesize (Sec.~\ref{SS:ControlMethod})};
			\node[format,right of=n2,draw = white,text width=1.75cm,xshift = 3.15cm](n2r){\footnotesize (Sec.~\ref{SS:grouping})};		
			\node[format,right of=n3,draw = white,text width=1.75cm,xshift = 3.15cm](n3r){\footnotesize (Sec.~\ref{SS:TUgame})};			
			\node[format,right of=n4,draw = white,text width=1.75cm,xshift = 3.15cm](n4r){\footnotesize (Sec.~\ref{SS:PayF})};	
		\end{tikzpicture}
	\end{center}
	\caption{Flowchart illustration of the proposed mechanism.}\label{F:FlowChart_mechanism}
\end{figure}
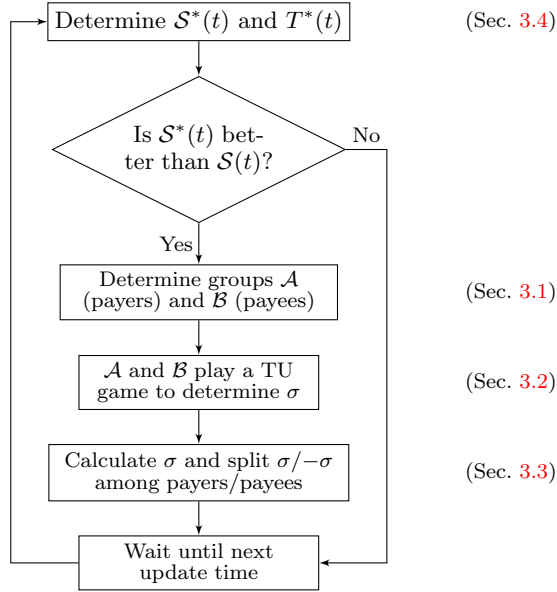

\label{S:DAlgorithm}
The flowchart of Fig.~\ref{F:FlowChart_mechanism} shows a simple framework of our intersection operation mechanism for an isolated junction. It also indicates the organization of this section (Sec.~\ref{S:DAlgorithm}). 
$S^*(t)$ and $T^*(t)$ are the new control strategy and its discharge vector at time $t$, $S(t)$ is the \emph{status quo} (or old) control strategy, $\A$ and $\B$ are the payer group (those that save time) and the payee group (those that incur additional travel time) in the new strategy, and $\sigma$ is the total payment from the payer group to the payee group.

\subsection{Vehicle Grouping}
\label{SS:grouping}
A control strategy can be thought of as discharge ordering for the vehicles at the intersection.  Therefore, a change in the control strategy $S(t) \mapsto S^*(t)$ affects the travel times of the vehicles at the intersection in one of three ways: their travel time would either increase, decrease, or remain unchanged under the new strategy (relative to the original one).  We group vehicles accordingly: those whose travel times would decrease under $S^*(t)$ (payers), those whose travel times would increase under $S^*(t)$ (payees), and those whose travel times are unaffected by the switch from $S(t)$ to $S^*(t)$ (indifferent vehicles). We also allow for vehicles not to be involved in the intersection game.  We set the VOT for these vehicles to zero, and they belong to the indifferent group. \textbf{Note}: This grouping is not to be confused with the intersection phases (movements that receive priority at time $t$ under the control strategy).  A vehicle in any of the three group may or may not be part of the signal phase that is to be served next under $S^*(t)$.

Suppose there are $n$ vehicles along the inbound links at the intersection (all approaches) at time $t$, and they are ordered in ascending order of arrival times,  $\{1, 2,\hdots, n\}$. Let time $t$ mark an instant at which the signal controller updates its strategy.  The update cadence can be uniform using a rolling horizon (e.g., every 10 seconds), triggered by a vehicle arrival (via communication between the vehicle and the controller), or a sensor actuation.  We denote by $\mathcal{S}(t)$ the control strategy at time $t$, prior to the update and denote the new strategy by $\mathcal{S}^*(t)$.  Let $\tau_{t,v}$ denote the discharge time of vehicle $v \in \{1,\hdots,n\}$ according to strategy $\mathcal{S}(t)$ and denote the associated vehicle discharge vector by $T(t)  \equiv [\tau_{t,1} \hdots  \tau_{t,n}]^{\top}$.  Similarly, the discharge vector under the updated control strategy is denoted by $T^*(t)  \equiv [\tau_{t,1}^* \hdots  \tau_{t,n}^*]^{\top}$.

Define the time difference vector $\Delta T_{t} \equiv T_{t} -T_{t}^*$, which will include non-negative elements \textbf{and} non-positive elements (due to possible re-ordering of vehicles in the discharge sequence).  Let $\Delta \tau_{t,v}$ denote the travel time gain of vehicle $v$ at time $t$; in other words, $\Delta \tau_{t,v}$ is the element of the vector $\Delta T(t)$ corresponding to vehicle $v$: $ \Delta T(t) = [\Delta \tau_{t,v} \hdots \Delta \tau_{t,n}]^{\top}$.

\begin{figure}[ht!] 
	\centering
	\subfigure[Discharge times under $\mathcal{S}(t)$]{\resizebox{0.33\textwidth}{!}{
			\includegraphics[width=0.35\textwidth]{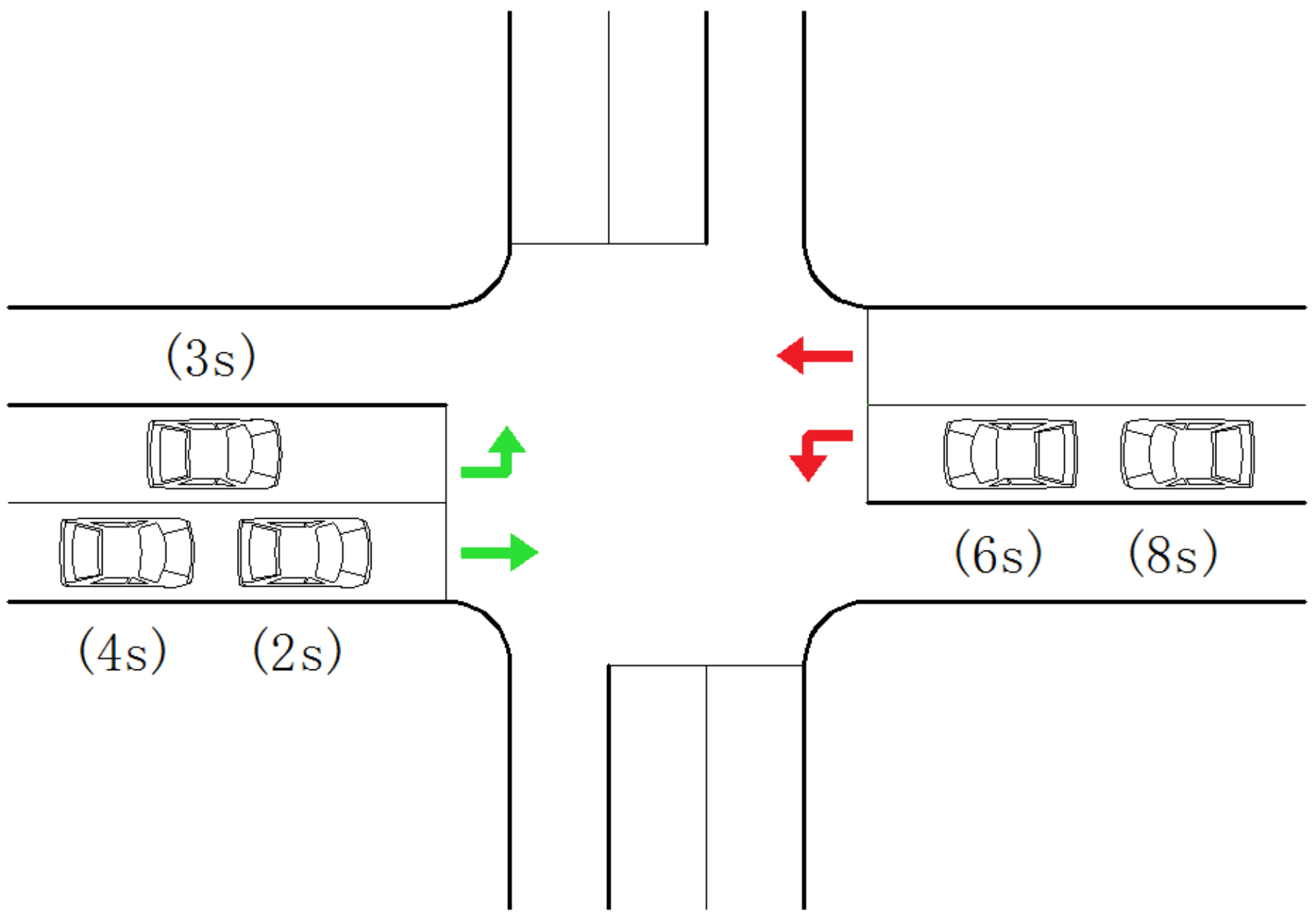}} 
		\label{F:Old_strategy}} 		
	
	\subfigure[Discharge times under $\mathcal{S}^*(t)$]{\resizebox{0.33\textwidth}{!}{
			\includegraphics[width=0.35\textwidth]{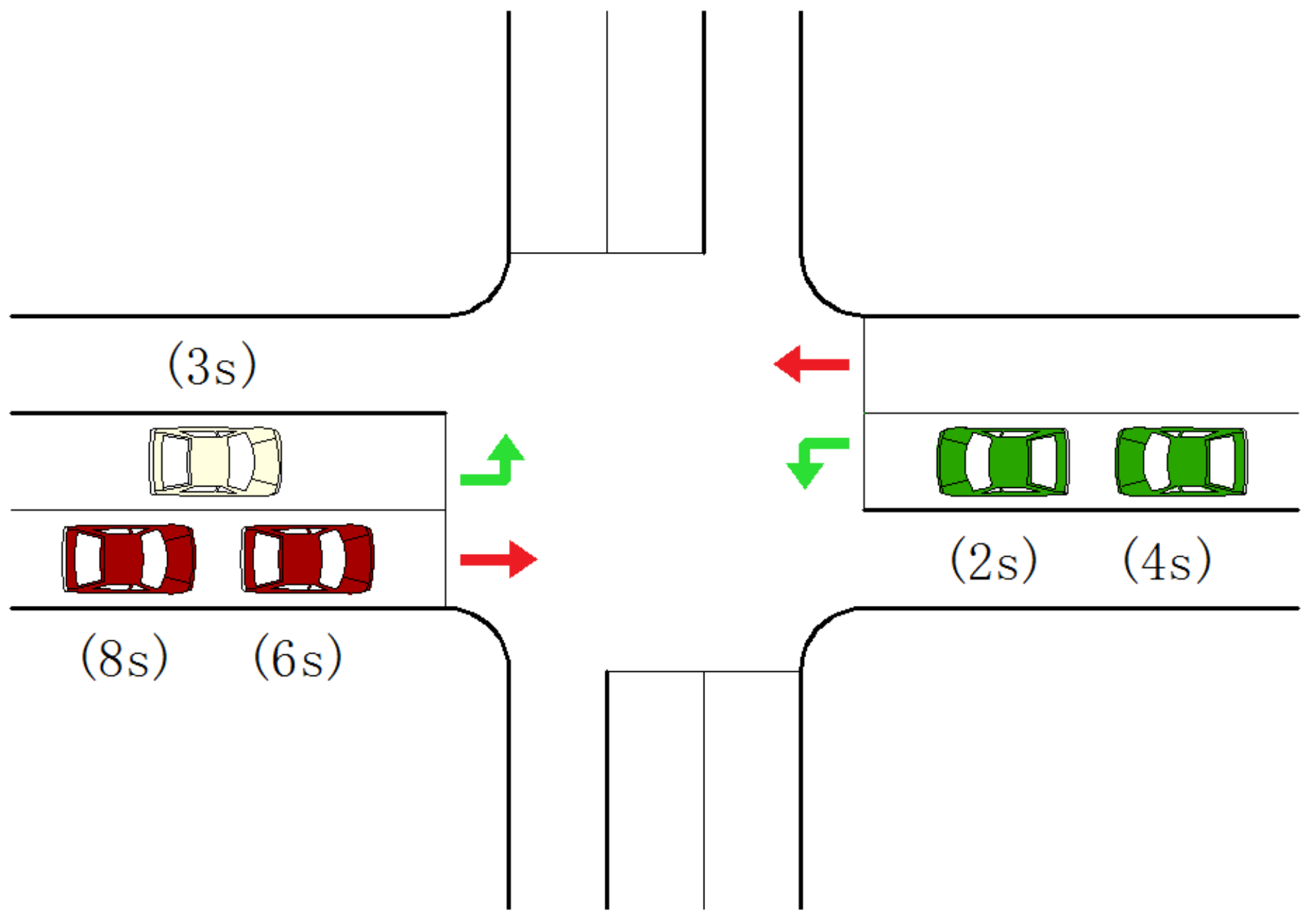}} 
		\label{F:New_strategy}}
	\caption{Example vehicle grouping, group $\A$ in green and group $\B$ in red.}
	\label{F:GameSample}
\end{figure}

We define the sets $\A$, $\B$ and $\C$ in accordance with whether their travel times improve (payers), worsen (payees), or remain the same (indifferent vehicles), respectively. We also explicitly exclude vehicles in the system that are not part of the game.  The latter are assumed to have zero VOT.  Let $\Upsilon_v$ denote the VOT of vehicle $v$, then
\begin{equation}
	\A \equiv \{v: \Upsilon_v \Delta \tau_{t,v} > 0\}
\end{equation}
\begin{equation}
	\B \equiv \{v: \Upsilon_v \Delta \tau_{t,v} < 0\}.
\end{equation}
and
\begin{equation}
	\C \equiv \{v: \Upsilon_v \Delta \tau_{t,v} = 0\}
\end{equation}
Since vehicles in $\C$ do not play the game and are not involved in transactions, the following analysis will only focus on groups $\A$ and $\B$. Note that $\A  \cap  \B = \emptyset$. Fig.~\ref{F:GameSample} illustrates how the two groups are selected.  The discharge times of the vehicles are enclosed in brackets.  Under the old control strategy in Fig.~\ref{F:GameSample} \subref{F:Old_strategy}, the eastbound through and left-turning vehicles are discharged first, followed by the westbound left-turning vehicles.  After the control update, the control strategy shown in Fig.~\ref{F:GameSample}\subref{F:New_strategy}, where the eastbound and westbound left-turning vehicles are discharged first, followed by the eastbound through vehicles.  For this update, there are two vehicles whose discharge times decrease (green), and two vehicles whose discharge times increase (red). 

\subsection{Transferable Utility (TU) Game Framework}
\label{SS:TUgame}
Our algorithm applies a cooperative game theory framework using a TU game similar to \cite{lin2019pay}. We describe this next.

\subsubsection{The TU game and side payments}
Consider an intersection game between two conflicting vehicle groups $\A$ and $\B$. We will envisage $\A$ and $\B$ as two players.   The payoff to $\A$ is $A_{mn}$ when action $(m,n) \in \{1,2\} \times \{1,2\}$ is chosen.  Similarly, $B_{mn}$ is the payoff to vehicle group $\B$ when action $(m,n) \in \{1,2\} \times \{1,2\}$ is chosen.  Here, we assume that action $m=1$ is preferred by vehicle group $\A$ and $n=1$ to be preferred by $\B$, while $m=2$ and $n=2$ are the ``disliked'' actions by $\A$ and $\B$, respectively.  We define 
\begin{equation}
	\mathbf{A} \equiv \begin{bmatrix} A_{11}&A_{12}\\A_{21}&A_{22}\\ \end{bmatrix} \mathrm{ and } ~ \mathbf{B} \equiv \begin{bmatrix} B_{11}&B_{12}\\B_{21}&B_{22}\\ \end{bmatrix}
\end{equation}
as the payoff matrices to $\A$ and $\B$, respectively. The TU game is a cooperative game, which allows side payment from one player to another.  

The feasible set for this TU game is the convex hull of the points $(A_{mn} - \sigma, B_{mn} + \sigma)$, where $\sigma \in \RR$ denotes the side payment.  We adopt the convention that $\sigma > 0$ corresponds to a side payment made from $\A$ to $\B$ while $\sigma<0$ represents a side payment made from $\B$ to $\A$.  Let $\omega_{mn}$ denote the total payoff that can be achieved by players $\A$ and $\B$ under action $(m,n)$, that is, $\omega_{mn} = (A_{mn} - \sigma) + (B_{mn} + \sigma) = A_{mn} + B_{mn}$.  Hence, in the presence of side payment, rationality dictates that the action chosen is that which maximizes the total payoff:
\begin{equation}
	(m^*,n^*) = \underset{(m,n) \in \{1,2\} \times \{1,2\}}{\arg \max} \big(A_{m,n} + B_{m,n} \big).
	\label{E:mn}
\end{equation}
Denote the maximal payoff by $\omega^*$, then
\begin{equation}
	\omega^* \equiv A_{m^*n^*} + B_{m^*n^*}.
	\label{E:omega}
\end{equation}
In the absence of agreement, $\A$ and $\B$ chose mixed strategies that ensure guaranteed payoffs independent of the other player's strategy.  This is known as the \textit{threat strategy}.  Let $\mathbf{p} \equiv [p \quad 1-p]^{\top}$ denote $\A$'s threat strategy and let $\mathbf{q} \equiv [q \quad 1-q]^{\top}$ denote $\B$'s threat strategy. Here $p$ is the probability that $\A$ chooses their preferred action $m=1$ and $q$ is the probability that $\B$ chooses their preferred action $n=1$. It follows that the expected payoff received by $\A$ under their threat strategy is $S_{\A} \equiv \mathbf{p}^{\top} \mathbf{Aq}$ and $\B$'s expected payoff under their threat strategy is $S_{\B} \equiv \mathbf{p}^{\top} \mathbf{Bq}$.  

The  solution to the TU game is one such that $\A$ accepts no less than $S_{\A}$ in expected payoff and $\B$ accepts no less than $S_{\B}$.  This defines two extreme points along the pareto-optimal frontier with payoff pairs $(S_{\A}, \omega^* - S_{\A})$ and $(\omega^* - S_{\B}, S_{\B})$.  Any convex combination of these two points is a TU solution and, particularly, the midway point constitutes a \textit{natural compromise} \cite{myerson2013game}.  The resulting payoff is $\frac{1}{2}(\omega^* + S_{\A} - S_{\B})$ to $\A$ and $\frac{1}{2}(\omega^* - S_{\A} + S_{\B})$ to $\B$.  Hence,
\begin{equation}
	A_{m^*n^*} - \sigma = \frac{\omega^* + S_{\A} - S_{\B}}{2}
\end{equation}
so that the side payment from $\A$ to $\B$ is given by
\begin{equation}
	\sigma = \frac{- \omega^* - S_{\A} + S_{\B}}{2} + A_{m^*n^*}. \label{E:A}
\end{equation}
Similarly, our analysis implies that
\begin{equation}
	B_{m^*n^*} + \sigma = \frac{\omega^* - S_{\A} + S_{\B}}{2}
\end{equation}
so that
\begin{equation}
	\sigma = \frac{\omega^* - S_{\A} + S_{\B}}{2} - B_{m^*n^*}. \label{E:B}
\end{equation}
Subtracting \eqref{E:A} from \eqref{E:B}, we get
\begin{equation}
	0 = \omega^* - A_{m^*n^*} - B_{m^*n^*}.
\end{equation}
which is true by definition, see \eqref{E:omega}.  This demonstrates consistency and provides justification for choosing the midway point (the natural compromise) solution to the TU game.

\subsubsection{Threat strategy}
To determine the threat strategies of $\A$ and $\B$, first note that since their average payoffs are $\frac{1}{2}(\omega^* + S_{\A} - S_{\B})$ to $\A$ and $\frac{1}{2}(\omega^* - S_{\A} + S_{\B})$ to $\B$, we have that $\A$ is naturally incentivized to select a threat strategy that maximizes $S_{\A} -S_{\B}$ while $\B$ seeks to minimize $S_{\A} - S_{\B}$.  Since, 
\begin{equation}
	S_{\A} - S_{\B} = \mathbf{p}^{\top} \mathbf{Aq} - \mathbf{p}^{\top} \mathbf{Bq} = \mathbf{p}^{\top}(\mathbf{A} - \mathbf{B}) \mathbf{q},
\end{equation}
finding the threat strategy can be described as a zero sum game with payoff matrix $\mathbf{A} - \mathbf{B}$.  If a saddle-point exists, the threat point corresponds to the saddle-point: the $(m,n)$ pair for which $\max_m \min_n (A_{mn} - B_{mn}) = \min_n \max_m (A_{mn} - B_{mn})$. Otherwise, the solution is simply an optimal two-person zero-sum game mixed strategy. In this game, $\A$ selects their strategy, $\mathbf{p}$, in a way that they are guaranteed the same (expected) payoff regardless of whether $\B$ chooses $n=1$ or $n=2$.  Similarly, $\B$ selects a strategy that guarantees the same (expected) payoff regardless of whether $\A$ chooses $m=1$ or $m=2$.  As such, $\A$ chooses a strategy $\mathbf{p}$ that renders $\B$ indifferent between choosing $n=1$ or $n=2$.  Under $\mathbf{p}$, player $\B$'s expected utility given $n=1$ (i.e., given they choose $n=1$ with probability $q=1$) is
\begin{multline}
	[p \quad 1-p] \begin{bmatrix} B_{11} - A_{11} & B_{12} - A_{12} \\ B_{21} - A_{21} & A_{22} - A_{22} \end{bmatrix} \begin{bmatrix} 1 \\ 0 \end{bmatrix}  \\
	= \big((B_{11} - A_{11}) - (B_{21} - A_{21}) \big)p + B_{21} - A_{21}. \label{E:B1}
\end{multline}
Their expected utility given $n=2$ (i.e, when $q= 0$) is
\begin{multline}
	[p \quad 1-p] \begin{bmatrix} B_{11} - A_{11} & B_{12} - A_{12} \\ B_{21} - A_{21} & B_{22} - A_{22} \end{bmatrix} \begin{bmatrix} 0 \\ 1 \end{bmatrix} \\
	= \big((B_{12} - A_{12}) - (B_{22} - A_{22}) \big)p + B_{22} - A_{22}. \label{E:B2}
\end{multline}
For player $\B$ to be indifferent between between these two choices, player $\A$'s threat strategy should render \eqref{E:B1} and \eqref{E:B2} equal.  We get,
\begin{align}
	p = \frac{A_{22} - B_{22} - A_{21} + B_{21}}{A_{11} - B_{11}  + A_{22} - B_{22} - A_{12} + B_{12} - A_{21} + B_{21} }. \label{E:p}
\end{align}

Similarly, player $\B$ chooses their threat strategy $\mathbf{q}$ in such a way that player $\A$ is indifferent between $m=1$ and $m=2$.  Under $\mathbf{q}$, player $\A$'s expected utility from given $m=1$ is
\begin{multline}
	[1 \quad 0] \begin{bmatrix} A_{11} - B_{11} & A_{12} - B_{12} \\ A_{21} - B_{21} & A_{22} - B_{22} \end{bmatrix} \begin{bmatrix} q \\ 1-q \end{bmatrix}  \\
	= \big((A_{11} - B_{11}) - (A_{12} - B_{12}) \big)q + A_{12} - B_{12} \label{E:A1}
\end{multline}
and player $\A$'s expected utility given $m=2$ is
\begin{multline}
	[0 \quad 1]  \begin{bmatrix} A_{11} - B_{11} & A_{12} - B_{12} \\ A_{21} - B_{21} & A_{22} - B_{22} \end{bmatrix} \begin{bmatrix} q \\ 1-q \end{bmatrix}  \\
	= \big((A_{21} - B_{21}) - (A_{22} - B_{22}) \big)q + A_{22} - B_{22}. \label{E:A2}
\end{multline}
For player $\A$ to be indifferent between choosing $m=1$ or $m=2$, player $\B$'s threat strategy should render \eqref{E:A1} and \eqref{E:A2} equal.  Hence,
\begin{align}
	q = \frac{A_{22} - B_{22} - A_{12} + B_{12}}{A_{11} - B_{11}  + A_{22} - B_{22} - A_{12} + B_{12} - A_{21} + B_{21}}. \label{E:q}
\end{align}
The threat point follows immediately.	In instances where the solutions \eqref{E:p} or \eqref{E:q} would return negative values, one may resort to the linear programming solution of the zero-sum mixed strategy game. Specifically, letting $C \equiv A - B$ denote the payoff matrix, one solves
\begin{equation}
	q = \underset{0 \le x \le 1}{\arg \min} \max \big\{ (C_{11} - C_{12})x + C_{12}, (C_{21} - C_{22})x + C_{22} \big\}
\end{equation}
using linear programming techniques.  It is well known that the dual problem produces the solution $p$ for player $\A$.

\subsection{Payoff Formulation}
\label{SS:PayF}
\subsubsection{Modeling Payoffs}
\label{SSS:payoff}
We define the gains to vehicle groups $\A$ and $\B$ associated with the updated strategy $\mathcal{S}^*(t)$ as
\begin{align}
	G_{\A} = \sum_{v \in \A} \Upsilon_v \Delta \tau_{t,v}
\end{align}
and
\begin{align}
	G_{\B} = \sum_{v \in \B} \Upsilon_v \Delta \tau_{t,v},
\end{align}
respectively. 
Note that $G_{\A} > 0$ and $G_{\B} < 0$.  As such, group $\A$ prefers the new control strategy $\mathcal{S}^*(t)$, while group $\B$ prefers the prior control strategy, $\mathcal{S}(t)$.  

Following the definitions in \ref{SS:TUgame}, we have the action pairs $(\mathcal{S}^*(t),\mathcal{S}(t))$, $(\mathcal{S}^*(t),\mathcal{S}^*(t))$, $(\mathcal{S}(t),\mathcal{S}(t))$, and $(\mathcal{S}(t),\mathcal{S}^*(t))$. We use the gains defined above as the payoffs to the vehicle groups when there is no conflict. The action pairs $(\mathcal{S}^*(t),\mathcal{S}(t))$ and  $(\mathcal{S}(t),\mathcal{S}^*(t))$ correspond to conflicts. In these two cases, we assume that a controller breaks the conflict by applying a randomized strategy: it selects $\mathcal{S}^*(t)$ with probability 0.5 and $\mathcal{S}(t)$ with probability 0.5.  Then, the expected payoffs for $(m,n)=(\mathcal{S}^*(t),\mathcal{S}(t))$ and $(m,n)=(\mathcal{S}(t),\mathcal{S}^*(t))$ are set to zero and we have that
\begin{equation}
	\mathbf{A} \equiv \begin{bmatrix} 0 & G_{\A}/2 \\ -G_{\A}/2 & 0\\ \end{bmatrix} \label{E:EA}
\end{equation}
and
\begin{equation}
	\mathbf{B} \equiv \begin{bmatrix} 0 & G_{\B}/2 \\ -G_{\B}/2 & 0 \\ \end{bmatrix}. \label{E:EB}
\end{equation}
Hence,
\begin{equation}
	\mathbf{A} - \mathbf{B} = \frac{1}{2}\begin{bmatrix} 0 & G_{\A} - G_{\B} \\ -G_{\A} + G_{\B} &0\\ \end{bmatrix}.
\end{equation}
For this matrix, it can be immediately demonstrated that a saddle-point exists, since $\max_m \min_n (A_{mn} - B_{mn}) = \min_n \max_m (A_{mn} - B_{mn}) = 0$, and the threat-point is the action pair $(m,n) = (1,1)$.

\subsubsection{Allocation of side payment}
\label{SSS:sidepay}
The total side payment from $\A$ to $\B$, $\sigma$, is immediately calculated using Eq. \eqref{E:A} or Eq.  \eqref{E:B}.  The side payment is divided between the vehicles in the game in proportion to their gains.  Let $\sigma_v$ denote the side payment made by vehicle $v$.  Then,
\begin{equation}
	\sigma_v = \begin{cases} \frac{1}{G_{\A}} \Upsilon_v \Delta \tau_{t,v} \sigma, & \mbox{ if } v \in \A \\ \frac{-1}{G_{\B}} \Upsilon_v \Delta \tau_{t,v} \sigma, & \mbox{ if } v \in \B \\
		\quad \quad 0, & \mbox{ if } v \not\in \A \cup \B \end{cases}. \label{E:deltaAB}
\end{equation}
According to \eqref{E:deltaAB}, because $\sigma > 0$, every vehicle in $\A$ makes a positive side payment, and every vehicle in $\B$ ``makes'' a negative side payment (i.e., receives a payment).  

\subsection{Control Method}
\label{SS:ControlMethod}
The proposed TU game framework can be combined with various control methods. The optimal strategy $S^*(t)$ can differ from one control method to another under the same inputs. In this section, we show how the proposed framework can be combined with three existing control techniques, a phase switching method and a reservation-based control method (for isolated intersections), and a max-weight based approach for a network-wide decentralized controller.

\subsubsection{Phase Switching Method}
\label{SSS:phase}

\begin{figure}[ht!] 
	\centering
	\subfigure[Intersection layout]{\resizebox{0.21\textwidth}{!}{
			\includegraphics[width=0.35\textwidth]{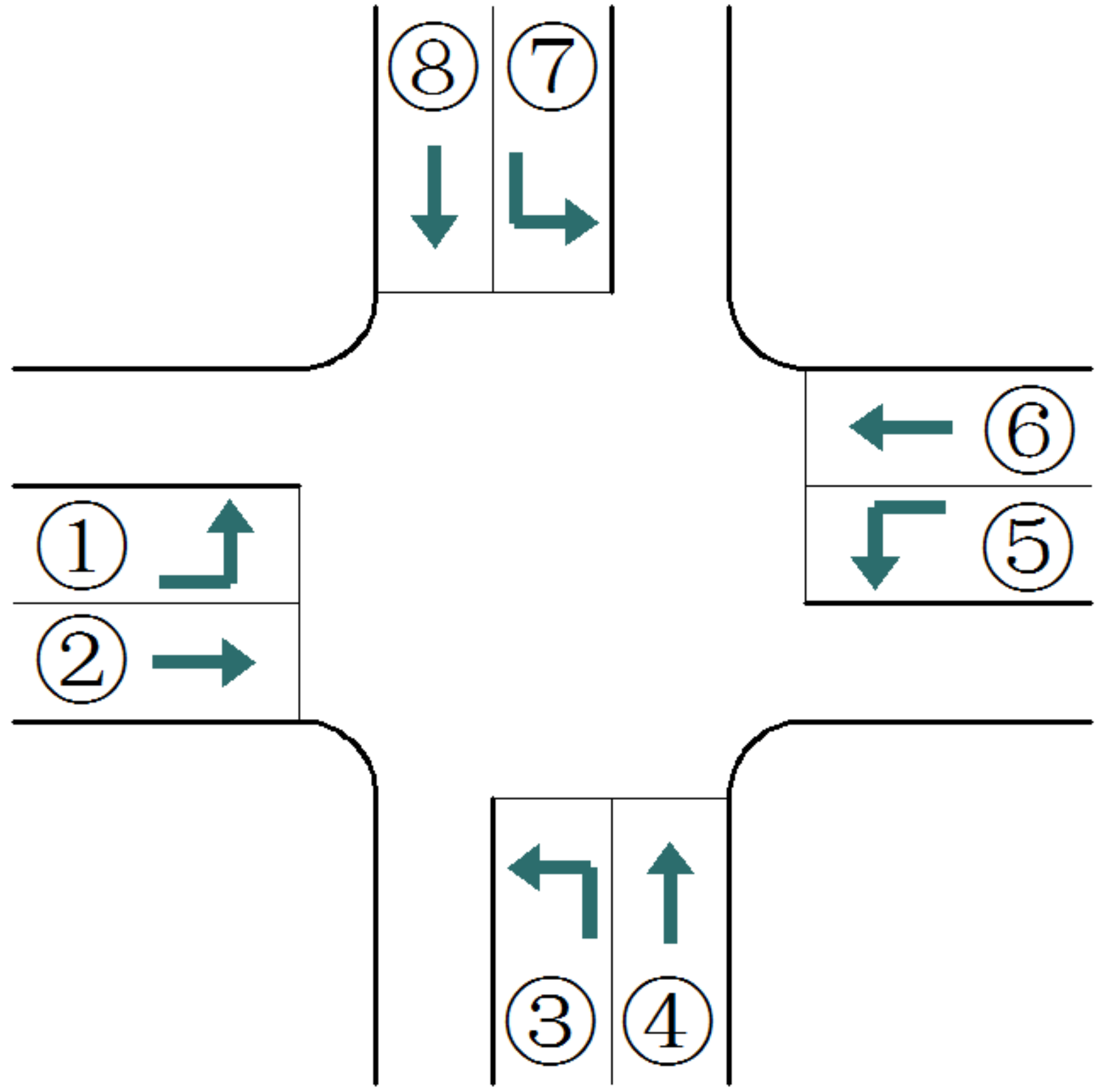}} 
		\label{F:Structure}} 	~~~
	\subfigure[Conflict-free signal phases]{\resizebox{0.23\textwidth}{!}{
			\includegraphics[width=0.35\textwidth]{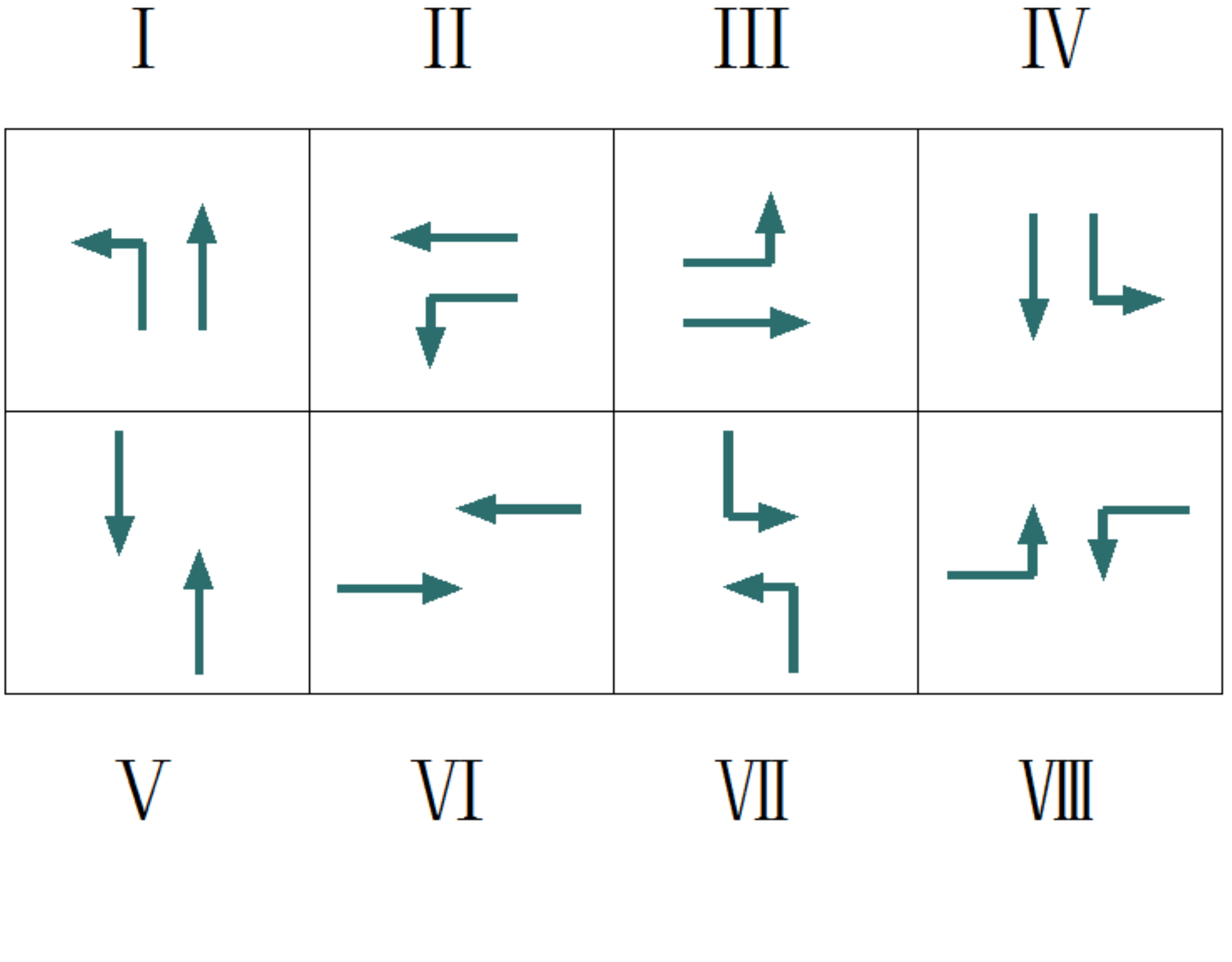}}
		\label{F:Phases}}
	\caption{Isolated intersection layout and conflict-free signal phases.}
	\label{F:Intersection}
\end{figure}

Without loss of generality, consider the example intersection layout in  Fig.~\ref{F:Intersection}\subref{F:Structure}, which is an isolated three-lane four-leg intersection with four through movements and four left-turning movements. For this intersection there are only eight possible movement combination, which are illustrated by the phasing schemes shown in Fig.~\ref{F:Intersection}\subref{F:Phases}. 

To ensure that all movements are serviced, the eight phases can be divided into two signal control groups: a signal cycle that includes phases I - IV and a signal cycle that includes phases V - VIII. The phases can be sequenced in numerous ways for each of these two cycles. The sequences can be pre-set, based on sensor actuation, or an adaptive controller.  Suppose the cycle length is fixed and let $P = \{\pi_1,\hdots,\pi_{|\mathcal{P}|}\}$ denote the set of permutations of phases that constitute a cycle, where $\pi_i$ represents one such sequence of phases (of pre-defined length).  The departure times and the travel time gains also depend on the phase sequence, $\pi$.  To capture this, we write the travel time gains as functions of $\pi$: $\Delta \tau_{t,v}(\pi)$.  At decision time $t$, the control strategy of phase switching method is simply one that selects a phase sequence as follows:
\begin{equation}
	\mathcal{S}^*(t) = \underset{\pi \in P}{\arg \max} \sum_{v = 1}^{n} \Upsilon_{v} \Delta \tau_{t,v}(\pi).
\end{equation}
This can be interpreted as a control strategy which maximizes the VOT-weighted travel time gains of all vehicles in the systems.  The method covers an entire cycle, but the updates can take place on a finer cadence, e.g., the control strategy can be re-evaluated at the end of each individual phase or even after a pre-set minimum green time.  This is similar to a control strategy with a rolling horizon. 

\subsubsection{Reservation-Based Control}
\label{SSS:reservation}
In a reservation-based system, CVs send requests to the controller upon entering the control zone of the intersection. Upon receiving a new request, the controller updates $S^*(t)$.  A similar approach was proposed in \cite{dresner2004multiagent} for AVs. A reservation-based system optimizes individual vehicle discharge times (and trajectories in the case of AVs). Suppose there are $n$ vehicles along the inbound links at the intersection (all approaches), and a new vehicle enters the system at time $t$. Here, $\mathcal{S}(t)$ is the optimized control strategy prior to the arrival of the new vehicle with the new vehicle processed last in the sequence, that is, it is assigned the index $n+1$ and the discharge times are $T(t) = [t_{t,1} \cdots t_{t,n} ~ t_{t,n+1}]^{\top}$. 

The updated strategy $\mathcal{S}^*(t)$ is one that selects a discharge sequence of the $n+1$ vehicles which maximizes the total VOT-weighted travel time gains. Let $\mathcal{P}(n+1) \subseteq 2^{n+1}$ denote the set of admissible permutations of the $n+1$ vehicles.  By \emph{admissible}, we mean to preclude sequences that involve impossible over-taking (for example).  Similar to the phase switching control example above, the updated strategy can be chosen as follows:
\begin{equation}
	\mathcal{S}^*(t) = \underset{\nu \in \mathcal{P}(n+1)}{\arg \max} \sum_{v = 1}^{n} \Upsilon_{v} \Delta \tau_{t,v}(\nu),
\end{equation}
where we have emphasized dependence of the travel time gains on the chosen permutation.  

Note that neither of these two methods scales to large networks computationally.  Computing optimal strategies is only possible if these control techniques are applied in a decentralized way.  This motivates the next control strategy, which is decentralized but comes with network-wide performance guarantees.

\subsubsection{Max-Weight Control}
\label{SSS:maxweight}
This approach can be seen as a generalization of the phase switching approach described above.  In this example, the updated strategy represents the next phase to select at each intersection in the network. Let $\mathcal{L}$ and $\mathcal{J}$ denote the sets of network links and network junctions, respectively. With slight notation abuse, let $\mathcal{P}_j = \{\pi_i^j\}$ denote the set of phases that can be selected at junction $j$, $\mathcal{M}_j$ the set of movements across junction $j$, and $\mathcal{M}$ the set of all network movements.  Each phase $\pi_i^j$ consists of a set of \emph{non-conflicting} movements in $\mathcal{M}_j$. 

Let $Q_{a,b}(t)$ denote the size of the vehicle queue along link $a$ that is destined to link $b$ and consider the Lyapunov function
\begin{equation}
	L(t,\boldsymbol{\pi}) \equiv \frac{1}{2} \sum_{(a,b) \in \mathcal{M}} [w_{a,b}(t,\pi^j)]^+ Q_{a,b}(t)^2, \label{E:L}
\end{equation}
where $[\cdot]^+ \equiv \max \{0,\cdot\}$, $\boldsymbol{\pi} = [\pi^1 \cdots \pi^j \cdots \pi^{|\mathcal{J}|}]$ is a vector of junction phases, and $w_{a,b}(t,\pi^j)$ is a movement weight. 
Let $\mathcal{V}_{a,b}(t)$ denote the set of vehicles in the queue at link $a$ destined to link $b$ at time $t$, then the movement weights are calculated as
\begin{equation}
	w_{a,b}(t,\pi^j) = \sum_{v \in \mathcal{V}_{a,b}(t)} \Upsilon_{v} \Delta \tau_{t,v}(\pi^j).
\end{equation}
Let $\phi_{a,b}(t,\pi^j)$ denote the vehicle flux from link $a$ to link $b$ if phase $\pi^j$ is active and define the weight
\begin{multline}
	W_{a,b}(\pi^j) \\ \equiv \Big[w_{a,b}(t,\pi^j)Q_{a,b}(t) - \sum_{c:(b,c) \in \mathcal{M}}w_{b,c}(t,\pi^j)r_{b,c}Q_{b,c}(t)\Big]^+,
\end{multline}
where $r_{b,c}$ is the fraction of vehicles in link $b$ that are destined to link $c$.  Finally, the max-weight strategy is calculated, for each $ j \in \mathcal{J}$, as
\begin{equation}
	\mathcal{S}_j^*(t) = \underset{\pi^j \in \mathcal{P}_j}{\arg \max} \sum_{(a,b) \in \mathcal{M}_j} W_{a,b}(\pi^j) \phi_{a,b}(t,\pi^j), \label{E:MW}
\end{equation}
where $\mathcal{S}_j^*(t)$ is the updated strategy for junction $j$. 
It can be demonstrated analytically that ensures traffic stability at the network level by applying Lyapunov drift techniques \cite{neely2005dynamic,neely2010stochastic}, using the Lyapunov function \eqref{E:L}.

\section{Numerical Experiments}
\label{S:Algorithm}
\subsection{General Settings} 
\label{SS:Se}
\begin{figure}[ht!] 
	\centering
	\resizebox{0.275\textwidth}{!}{%
		\includegraphics{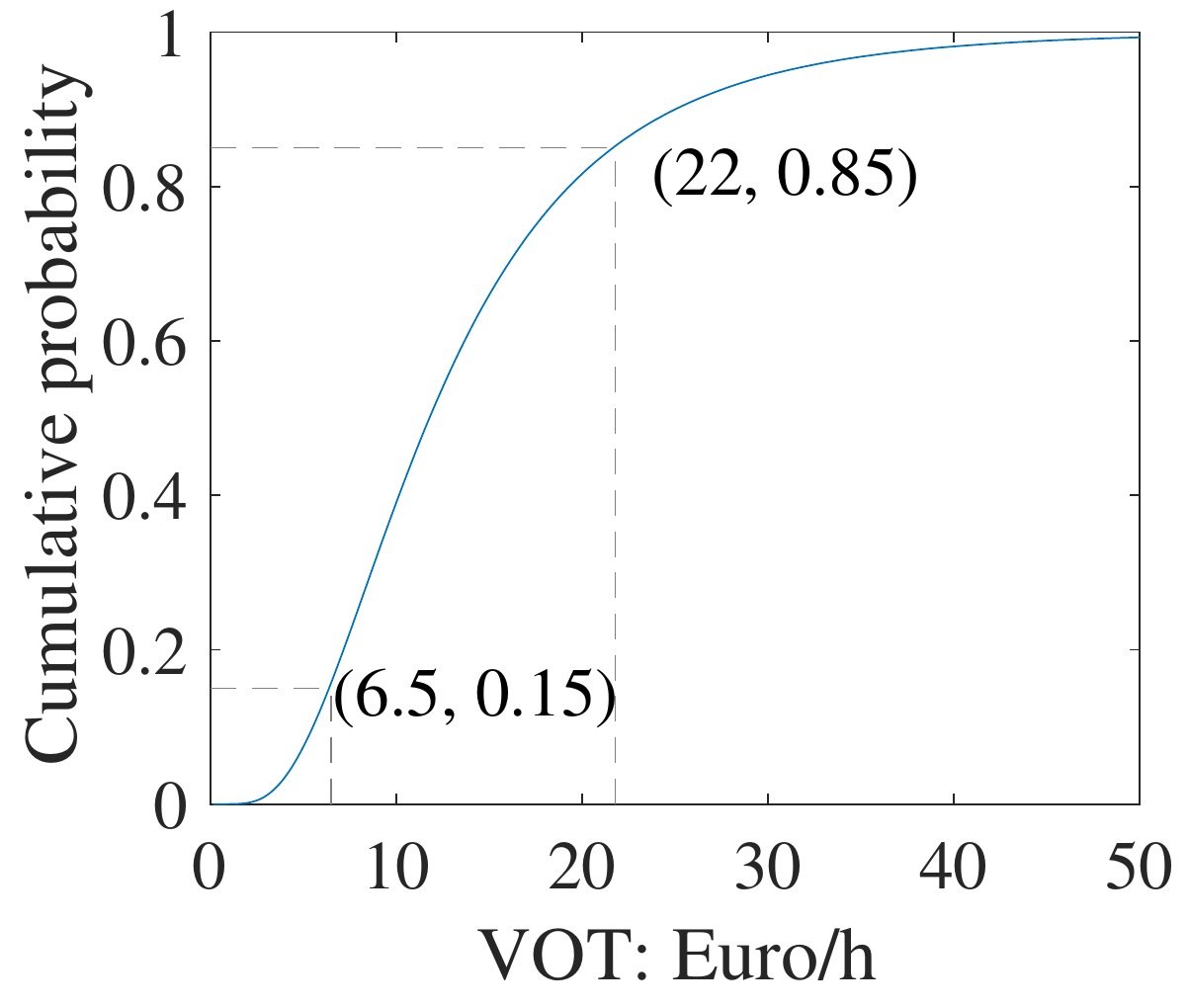}}	
	\caption{Cumulative distribution of VOT.}
	\label{F:VOT}%
\end{figure}

In the simulation experiments below, the minimum time headway is set to 1.8 seconds and the free-flow speed is 60 km/h. We use a log-normal distribution to generate vehicle VOTs, and assume a mean VOT of $14.1 \mathrm{Euro/h}$ \cite{asensio2008commuters} and a standard deviation of $9 \mathrm{Euro/h}$. The cumulative distribution function of VOTs is shown in Fig.~\ref{F:VOT}.
Vehicle inter-arrival times follow a negative exponential distribution with different rates for different experiments. The control zone of the intersections in the experiments are circles with different radii in different experiments, including 75m and 150m, corresponding to 10 vehicles/lane and 20 vehicles/lane at most.  These distances are well within the communication ranges of V2I \cite{maitipe2012development}. The control zone of the intersection can include both queued vehicles and vehicles that are in motion and approaching the intersection. For experiments with large arrival rates, we allow vehicles to queue outside of the intersection's control zone, such vehicles are not considered in the calculations (they are treated as being outside of the system from the controller's perspective). In each simulation scenario (fixed volume, control zone, penetration rate, etc.), we simulate at least 54,000 vehicle arrivals for a single intersection.

To analyze the performance of the proposed technique, we utilize a ``benefit'' metric denoted $\beta_v$, which is a quasi-linear function of time-saved and payment similar to \cite{vickrey1961counterspeculation}. We benchmark against a first-come-first-serve plan: we define the \textit{time saved} by vehicle $v$ as their travel time under a first-come-first-serve plan less their travel time using the proposed approach.  Denote the time saved by vehicle $v$ by $\tau_v^{\mathrm{s}} \in \RR$. Vehicle $v$'s benefit is given by
\begin{equation}
	\label{E:bv}
	\beta_v = \tau_v^{\mathrm{s}} \Upsilon_v - \sigma_v. 
\end{equation}
We also define another metric based on delay, which we refer to as loss. Denote by $\lambda_v$ the loss to vehicle $v$.  This is given by
\begin{equation}
	\label{E:lv}
	\lambda_v = \tau_v^{\dd} \Upsilon_v + \sigma_v,
\end{equation}
where $\tau_v^{\dd} \ge 0$ is vehicle $v$'s delay, defined here as their travel time less their free-flow (unobstructed) travel time.  

\subsection{Isolated Intersection Experiments} 
\label{SS:PI}
\begin{figure}[!ht] 
	\centering
	\subfigure[Low volume case]{\resizebox{0.22\textwidth}{!}{
			\includegraphics{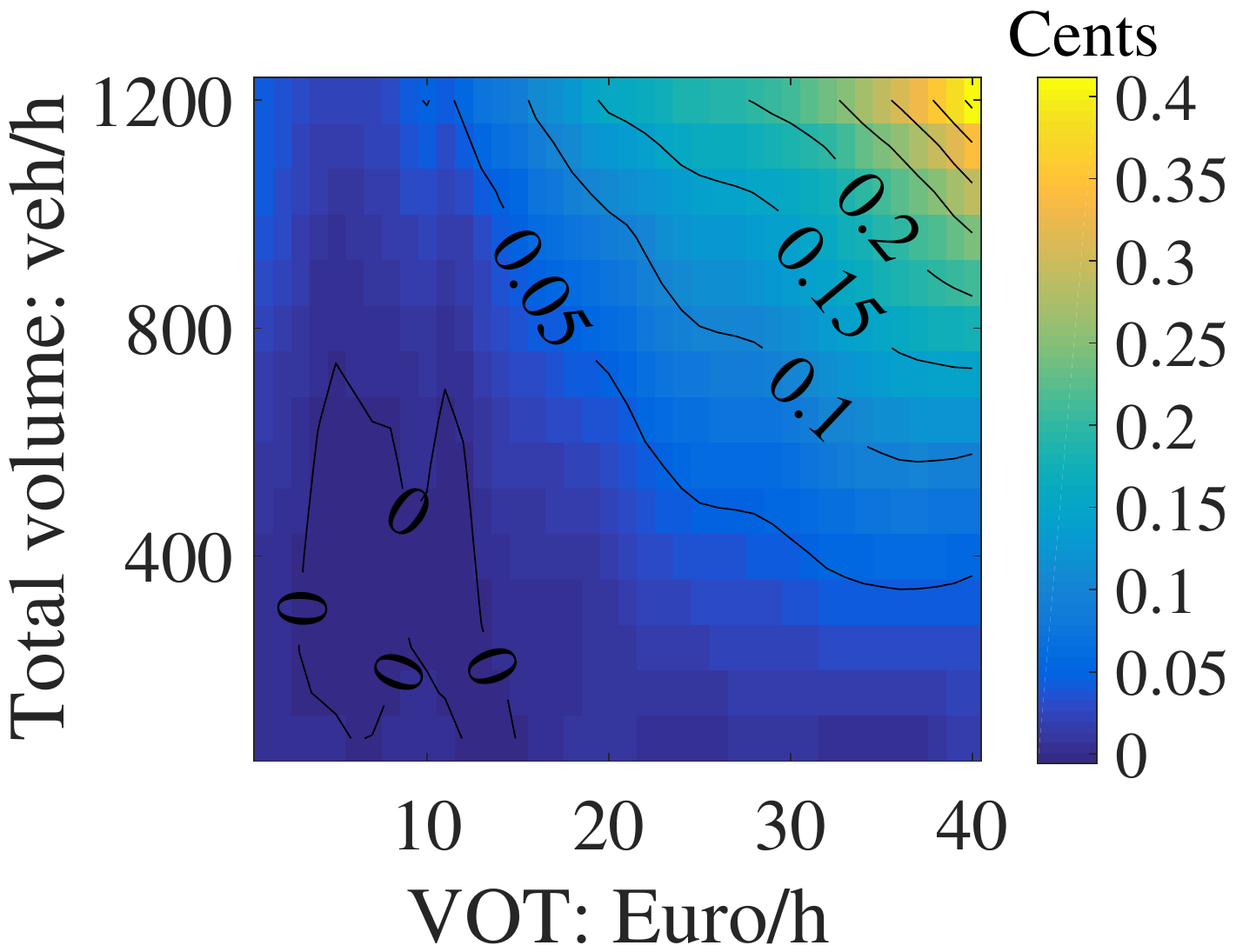}} 
		\label{F:B_small}} 
	\subfigure[High volume case]{\resizebox{0.21\textwidth}{!}{
			\includegraphics{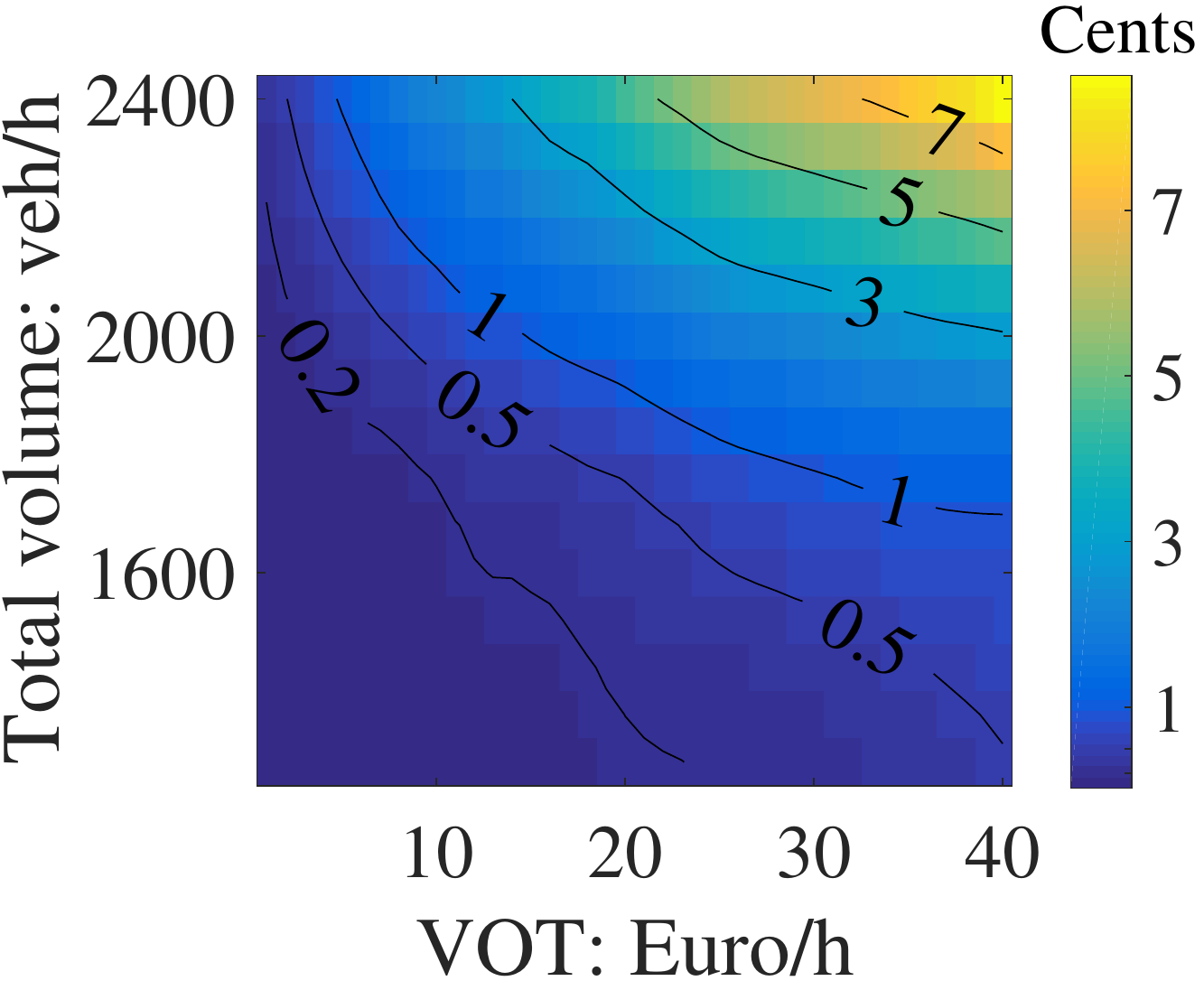}} 
		\label{F:B_high}}
	\caption{Contour of mean benefit under direct-transactions (baseline: first-come-first-serve).}
	\label{F:BF}
\end{figure}

To test the proposed approach, we perform simulation experiments on a hypothetical three-lane four-leg intersection (shown in Fig.~\ref{F:Intersection}). We use reservation-based control described in Sec.~\ref{SSS:reservation}.  Fig.~\ref{F:BF} shows the average benefit (mean of $\beta_v$ in \eqref{E:bv}, in cents) of our method from a vehicle's entry to its exit, in which the control zone is 20 vehicles/lane (150m), and penetration rate is 100\%.
Fig.~\ref{F:BF}\subref{F:B_small} summarizes low traffic volume scenarios (from 0 to 1200 veh/h), and Fig.~\ref{F:BF}\subref{F:B_high} summarizes high traffic volume scenarios (1200 to 2400 veh/h). 

\begin{figure}[ht!] 
	\centering
	\subfigure[Low volume case]{\resizebox{0.23\textwidth}{!}{
			\includegraphics[width=0.35\textwidth]{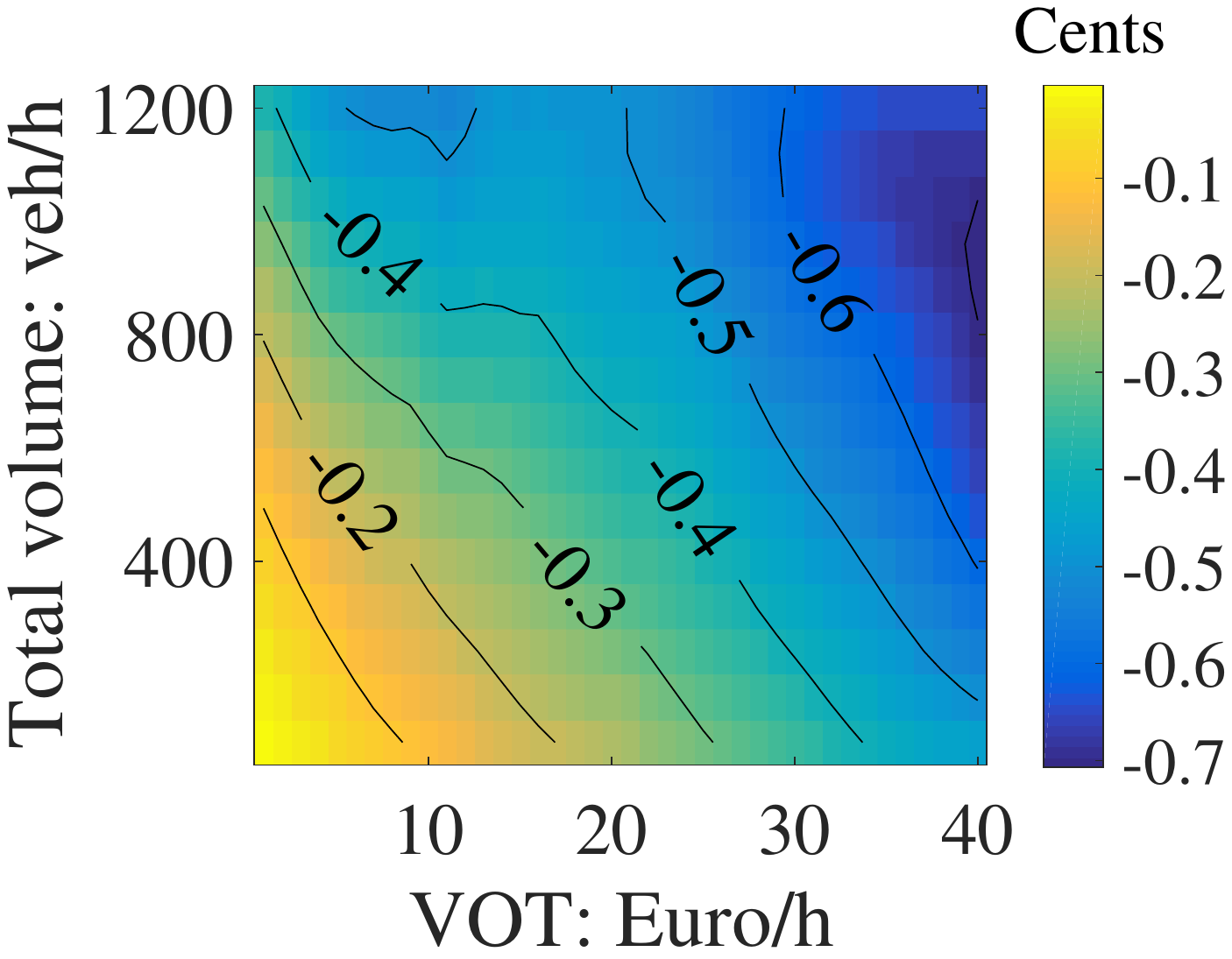}} 
		\label{F:B_Auction_small}} 		
	\subfigure[High volume case]{\resizebox{0.22\textwidth}{!}{
			\includegraphics[width=0.35\textwidth]{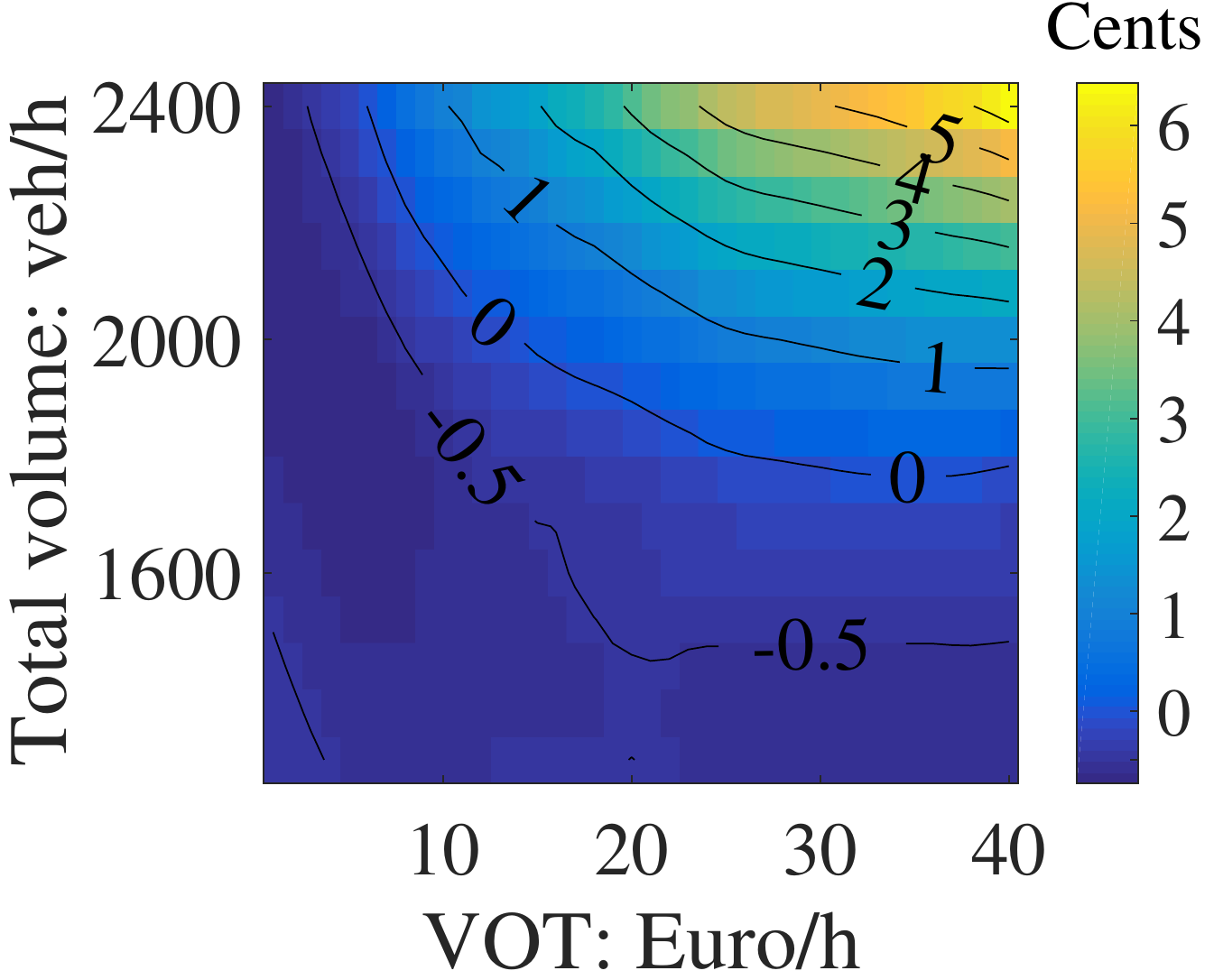}} 
		\label{F:B_Auction_high}}
	\caption{Contour of mean benefit under second-price auctions (baseline: first-come-first-serve).}
	\label{F:BF_Auction}
\end{figure}

In general, most vehicles have positive benefit on average. Furthermore, when VOT is higher, the benefit is larger; and when the traffic volumes increase, the benefits also increase rapidly. For comparison, using the same random seeds, we run simulation experiments with a second price auction scheme.  The results are shown in Fig.~\ref{F:BF_Auction}. Because winners are required to pay an operator, their expected benefit is always negative when the intersection volume is low (Fig.~\ref{F:BF_Auction}\subref{F:B_Auction_small}).
However, when the volume is high (Fig.~\ref{F:BF_Auction}\subref{F:B_Auction_high}), second-price auctions outperform first-come-first serve, since the latter do not account for VOT.  If we compare Fig.~\ref{F:BF} with Fig.~ \ref{F:BF_Auction}, we see that the mean benefit of direct-transactions is always higher than the mean benefit under second price auctions.

\begin{figure}[h!] 
	\centering
	\resizebox{0.28\textwidth}{!}{%
		\includegraphics{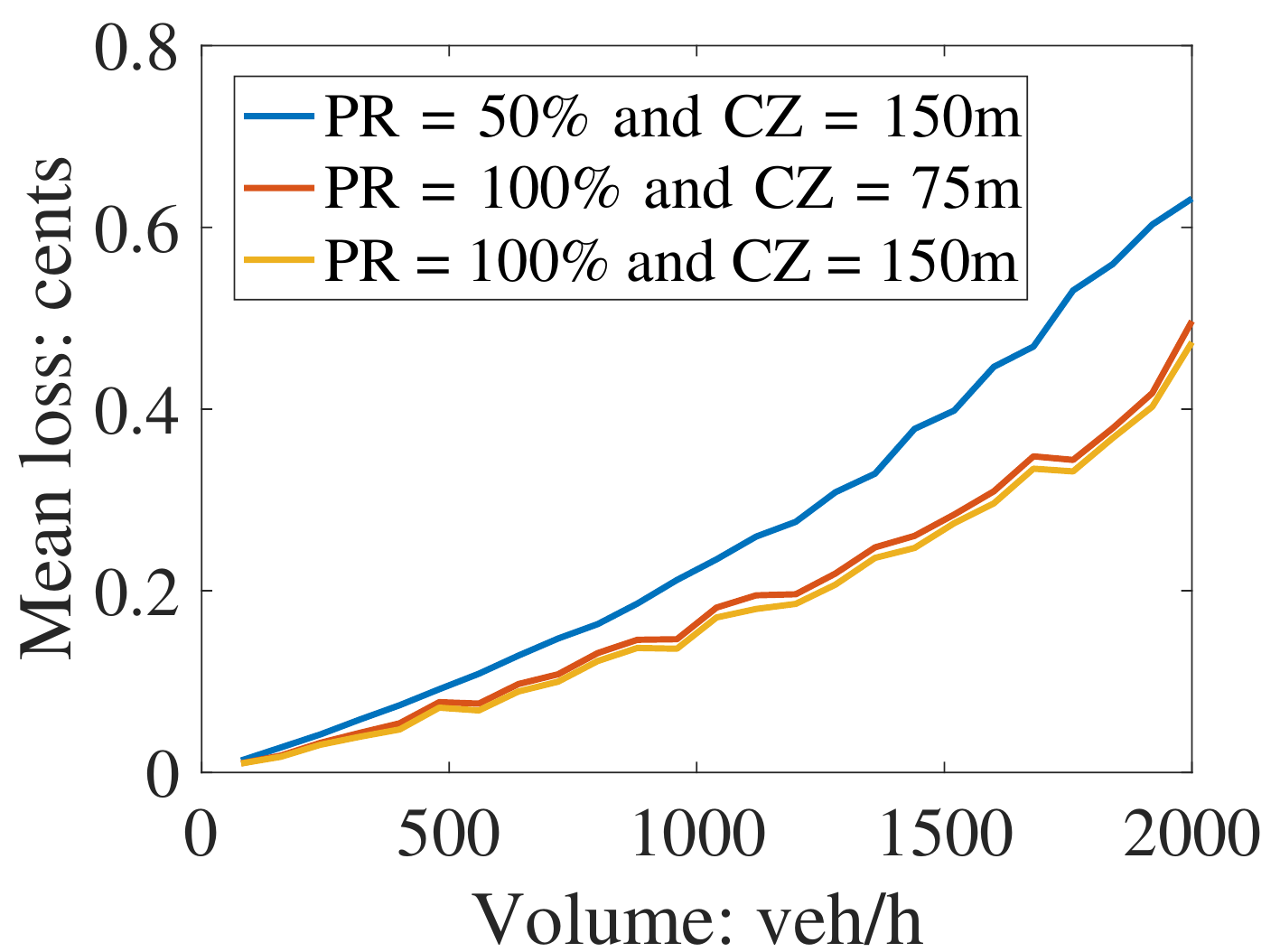}}	
	\caption{Sensitivity analysis of penetration rate and detection ability.}
	\label{F:sen_analysis}%
\end{figure}

Fig.~\ref{F:sen_analysis} provides a sensitivity analysis of the influence of penetration rate (PR) and control zone size (CZ). It shows that the mean loss (mean of $\lambda_v$ in \eqref{E:lv}) increases significantly with decreasing penetration rates, and it increases slightly with decreasing control zone size.

\begin{figure}[h!] 
	\centering
	\resizebox{0.28\textwidth}{!}{%
		\includegraphics{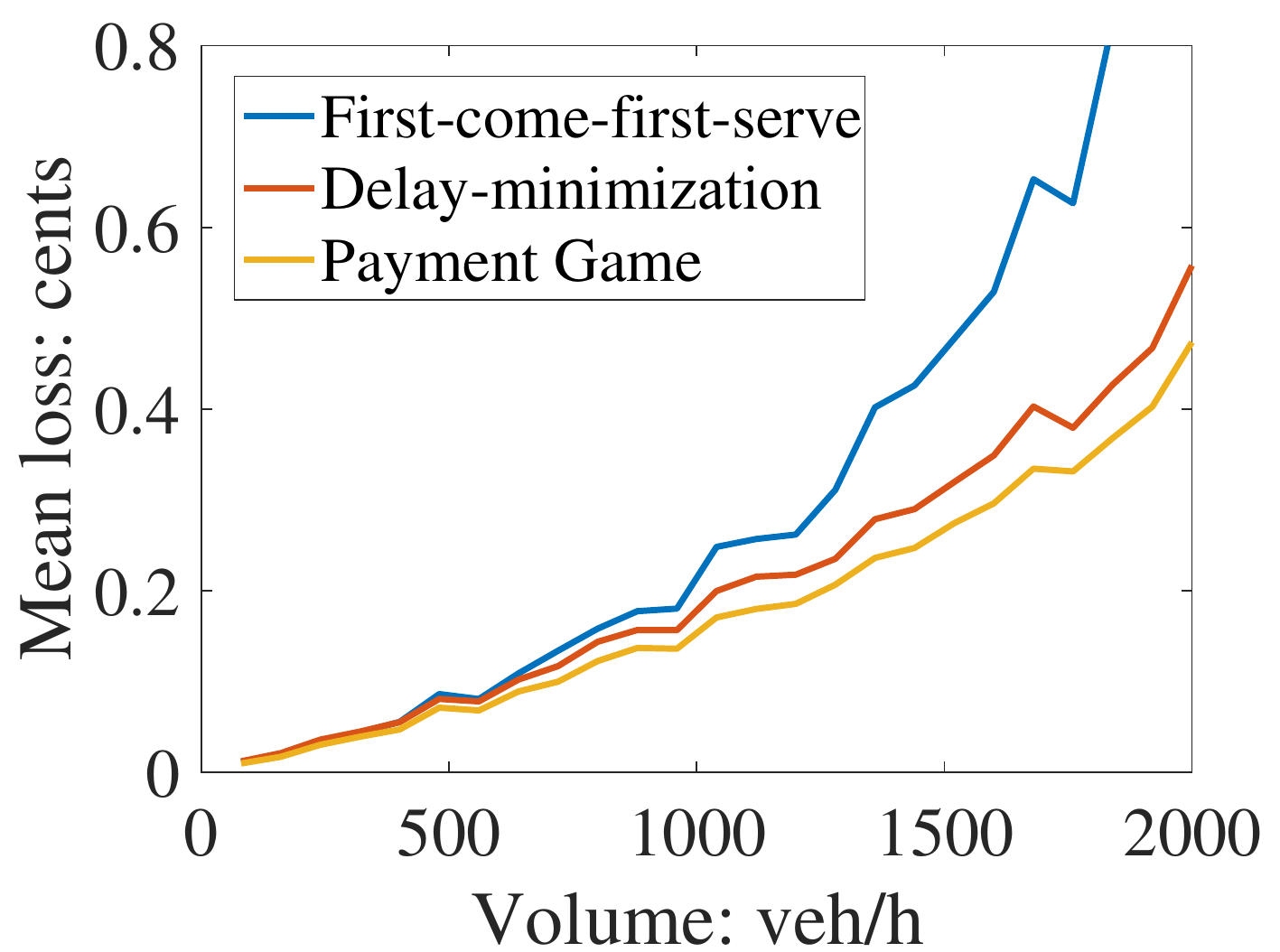}}	
	\caption{Performance of three control strategies.}
	\label{F:iso_comparison}%
\end{figure}

We next compare our results with two approaches that do not consider VOT or side payments under the same simulation framework. The first minimizes total delay and the second applies a first-come-first-served service discipline. The results are depicted in Fig.~\ref{F:iso_comparison}. The figure illustrates the competitiveness of the proposed approach as it outperforms the other two strategies in terms of mean loss averaged over all vehicles in the system.

\subsection{Impact of Untruthful VOT Reporting}
\label{SS:SimulationResults}
If a vehicle reports a false VOT (while others report true VOTs), the benefit (or loss) can change. The change in benefit as a result of this ``marginal'' dishonest reporting is given by
\begin{multline}
	\Delta \beta_v \equiv \beta_v^r  - \beta_v \\
	= [\tau_v^{\mathrm{s,false}} -\tau_v^{\mathrm{s,honest}}] \Upsilon_v - [\sigma_v^{\mathrm{false}} - \sigma_v^{\mathrm{honest}}], 
	\label{E:additional}
\end{multline}
where $\beta_v^r$ represents benefit with false VOT reporting, $\tau_v^{\mathrm{s,false}}$ is the time saved with false reporting, $\tau_v^{\mathrm{s,honest}}$ is the time saved with honest reporting, $\sigma_v^{\mathrm{false}}$ is the side payment made under false reporting, and $\sigma_v^{\mathrm{honest}}$ is the side payment that is made under honest reporting.  

\begin{figure}[h!] 
	\centering
	\resizebox{0.33\textwidth}{!}{%
		\includegraphics{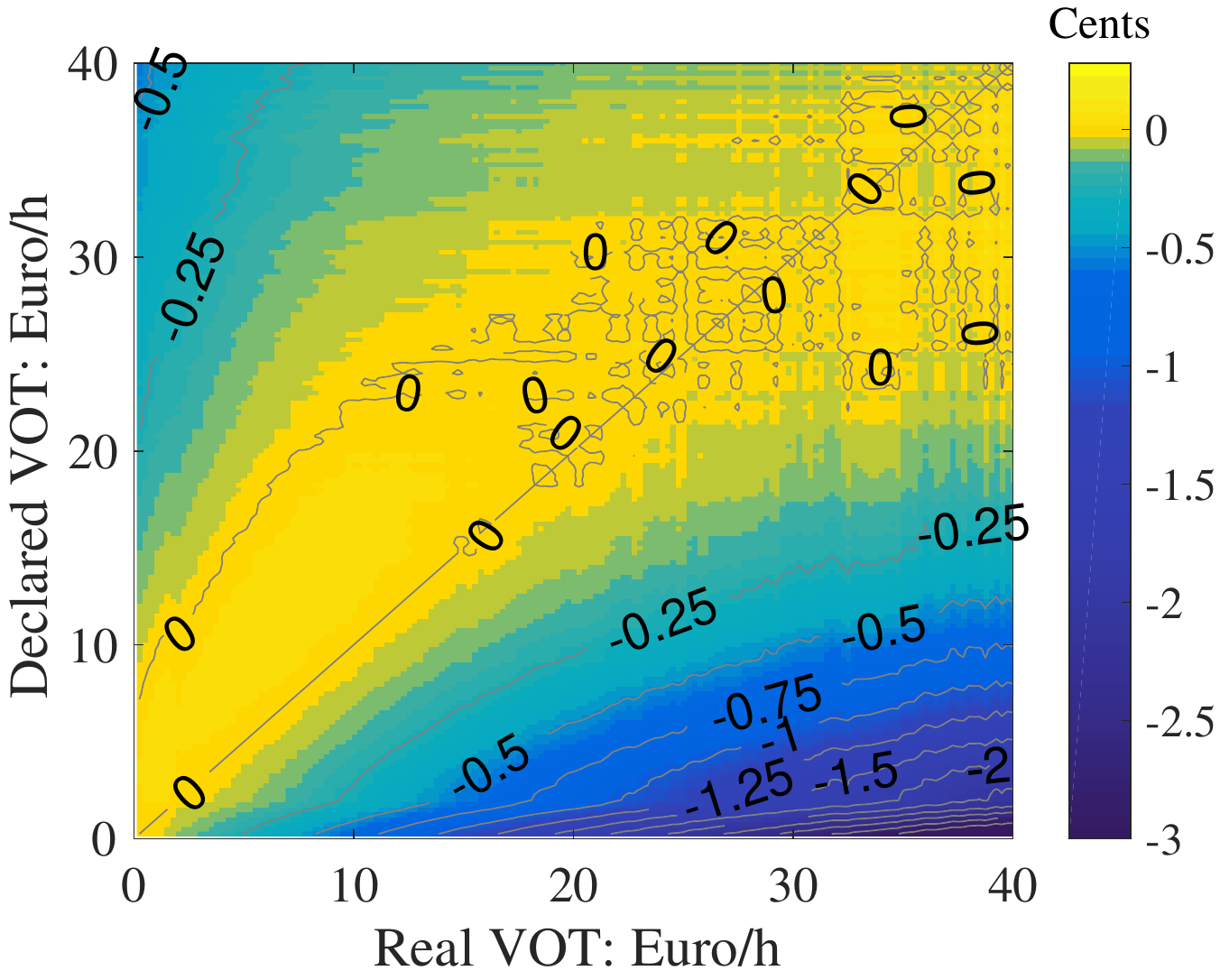}}	
	\caption{Additional benefit after untruthful reporting of VOT.}
	\label{F:lying}%
\end{figure}

On the three-lane four-leg intersection, we test the impact of untruthful VOT reporting. Fig.~\ref{F:lying} is a contour plot of $\Delta \beta_v$ values associated with the different combinations of true VOT and declared VOT, when the average traffic volume is 1200 veh/h.

The figure shows that the zero additional benefit lies along the diagonal.  This corresponds to the case where vehicles are truthful.  As we move away from the diagonal line, the benefit differences tend to be negative, which indicates that untruthfulness in VOT reporting should be dis-incentivized in the proposed framework.

\begin{figure}[h!] 
	\centering
	\resizebox{0.36\textwidth}{!}{%
		\includegraphics{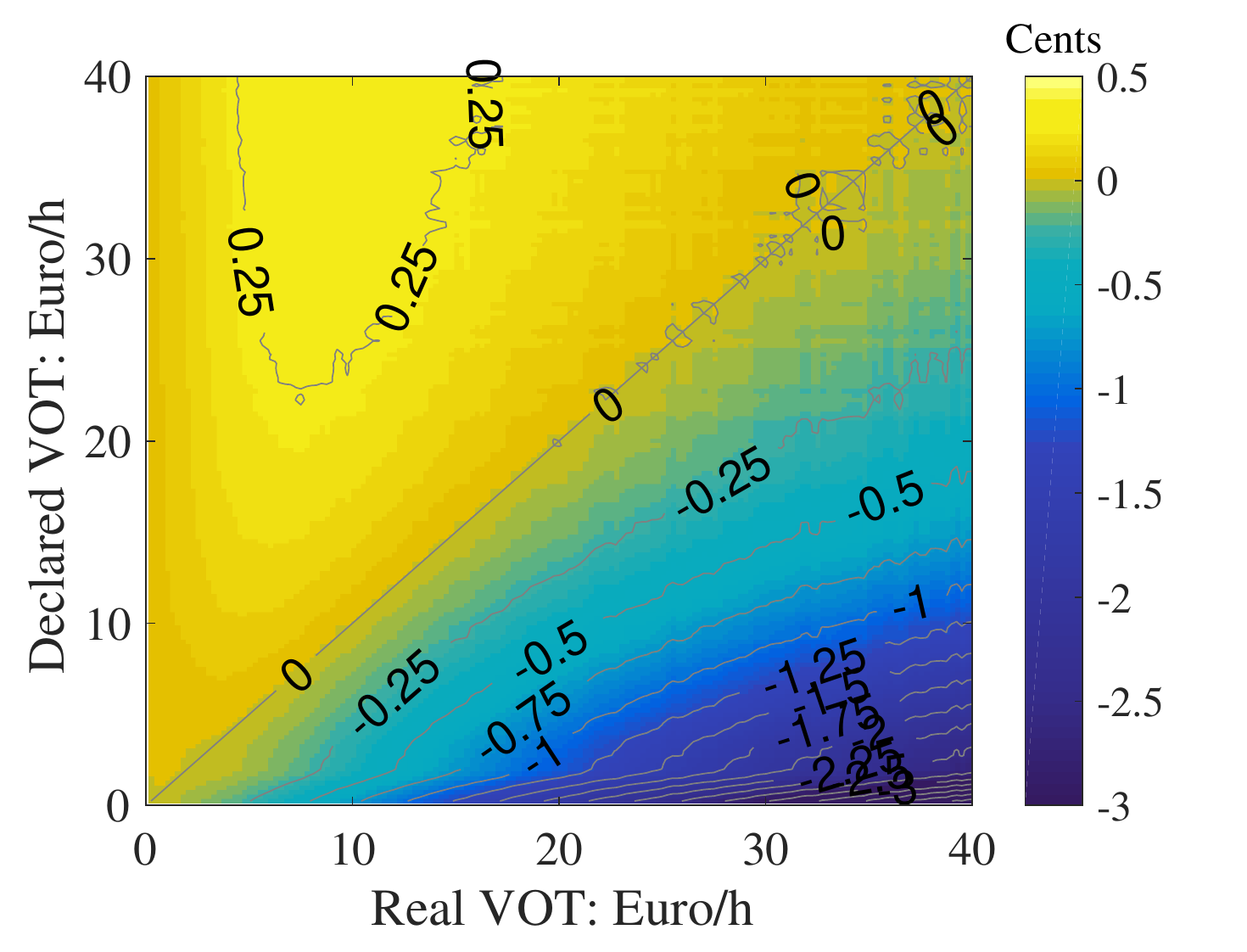}}	
	\caption{Additional benefit after untruthful reporting of VOT (without transaction).}
	\label{F:lying_no_income}%
\end{figure}

Next, consider the same experiment but without side payments. The net benefits, $\Delta \beta_v$, in this case are depicted in Fig.~\ref{F:lying_no_income}. The results suggest that there is always an incentive to exaggerate one's VOT. We see that basing priority on VOT but prohibiting side payments results in poor performance overall especially when the real VOTs get higher.

\subsection{`Gaming' the System}
In this experiment we simulate the following scenario: a vehicle that stops at an intersection approach and collects payments in exchange for providing priority to vehicles in an opposing approach thereby blocking their approach. The question is: how much benefit can this vehicle earn from such a strategy?

\begin{figure}[h!] 
	\centering
	\resizebox{0.3\textwidth}{!}{
		\includegraphics{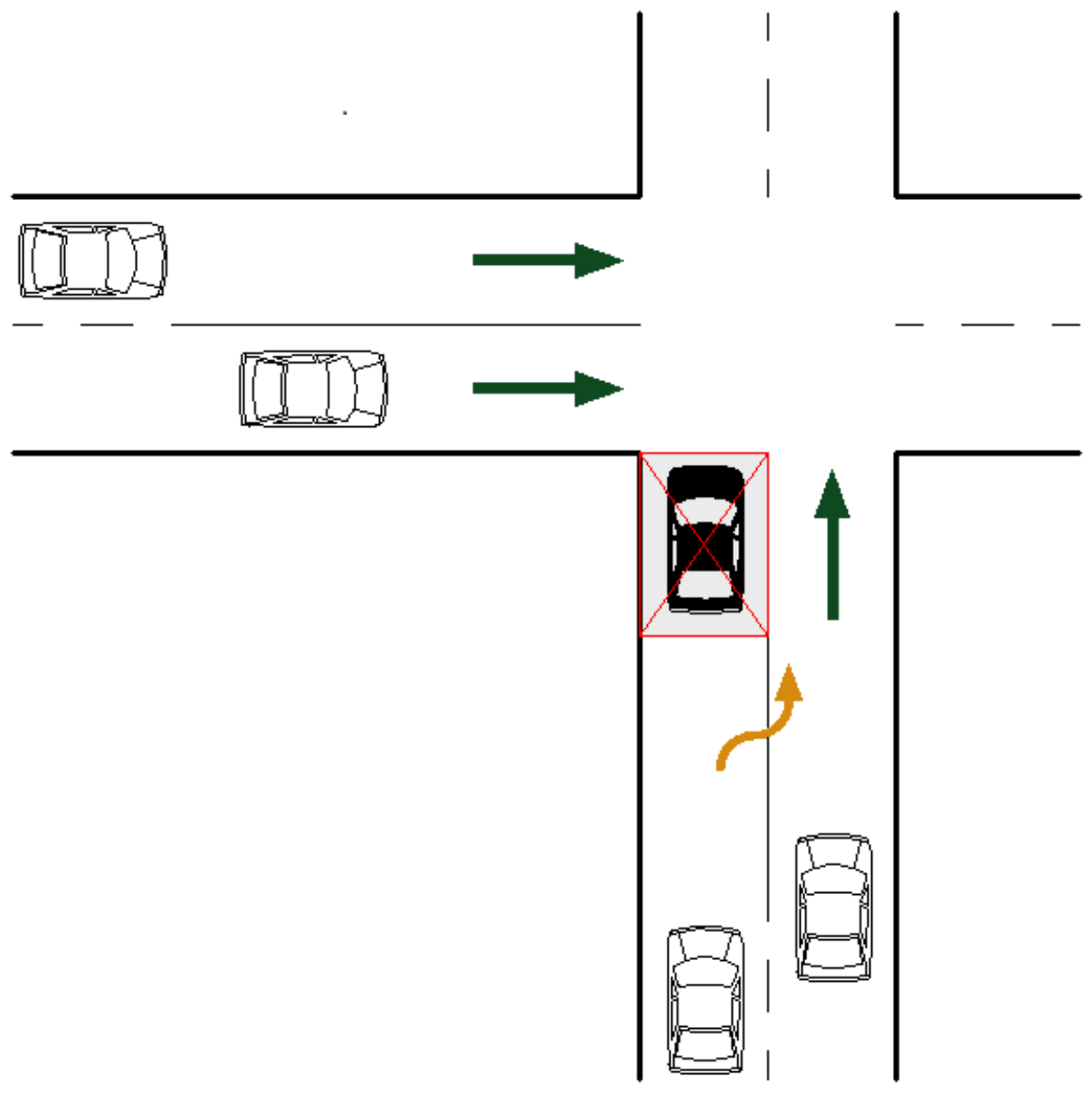}}	
	\caption{'Gaming' the system to generate income.}
	\label{F:Intersection_two}%
\end{figure}

In order to make the example more realistic, we set each movement to have two lanes so that one stopped vehicle will not block the approach completely. As shown in  Fig.~\ref{F:Intersection_two}, the black vehicle stops and blocks its lane, and its following vehicles need to change lanes to pass. We only take two movements and choose phase switching control in this toy sample. Vehicles in the blocked lane have higher headways and lower throughputs (less opportunity to change lane and pass) than their neighbors.

\begin{figure}[h!] 
	\centering
	\resizebox{0.3\textwidth}{!}{%
		\includegraphics{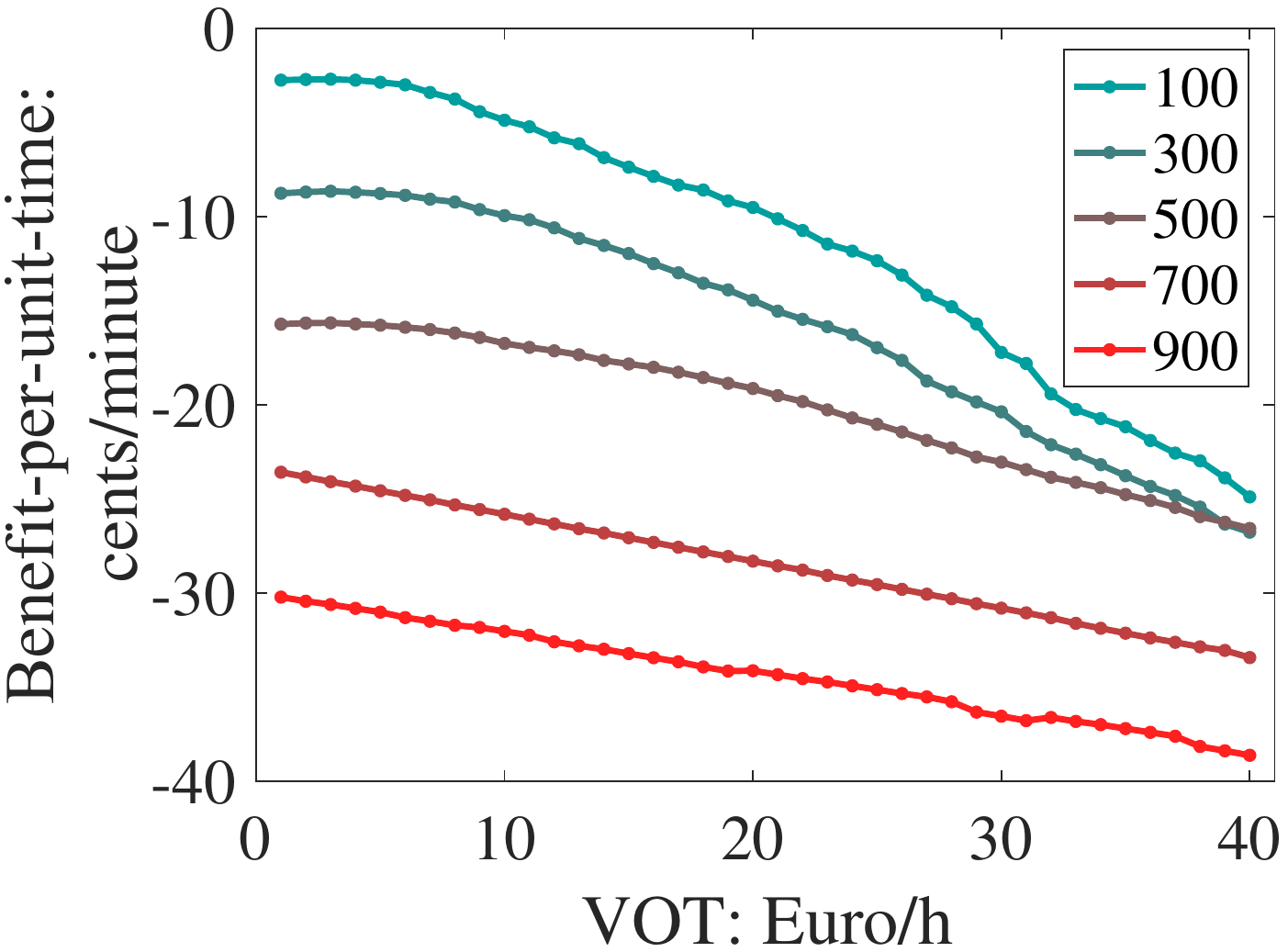}}	
	\caption{Benefit-per-minute of the obstructing vehicle.}
	\label{F:Benefit_violator}%
\end{figure}

We test with VOTs in the range 1-40 \$/h (in increments of 1\$/h) for the obstructing vehicle and various vehicle arrival rates to the system, from 100 to 900 veh/h/lane. The results, shown in Fig.~\ref{F:Benefit_violator}, illustrate a 'frustrating' situation for this vehicle. Because there is no 'time saved' for the obstructing vehicle, we redefine its benefit-per-unit-time as valuated time-wasted (1 minute times VOT) plus monetary income. We see that the benefit-per-unit-time in all scenarios is negative.  Moreover, we see that this loss increases with VOT and volume.

The primary reason for this is that playing the game is voluntary but if one chooses to play there is always a chance that they win and will hence be required to make a payment. In essence, the stopped vehicle delays the whole movement, which results in an accumulation of vehicles along the approach creating an asymmetry at the intersection.  This, in turn, results in the obstructing vehicle's approach winning the game more frequently. 

\subsection{Arterial Experiments} 
\label{SS:AI}
\begin{figure}[h!] 
	\centering
	\resizebox{0.45\textwidth}{!}{%
		\includegraphics{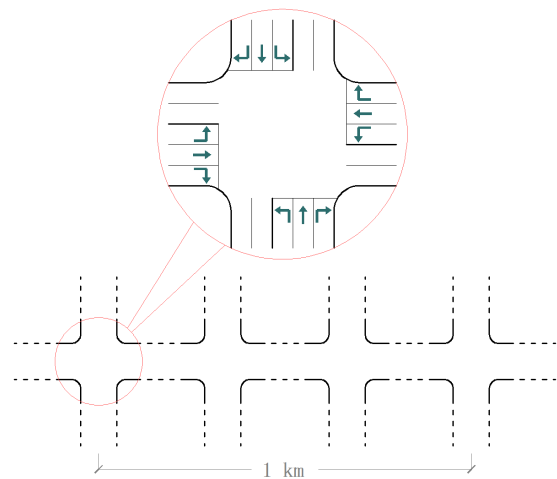}}	
	\caption{Arterial layout.}
	\label{F:Arterial}%
\end{figure}

\begin{figure}[h!] 
	\centering
	\resizebox{0.27\textwidth}{!}{%
		\includegraphics{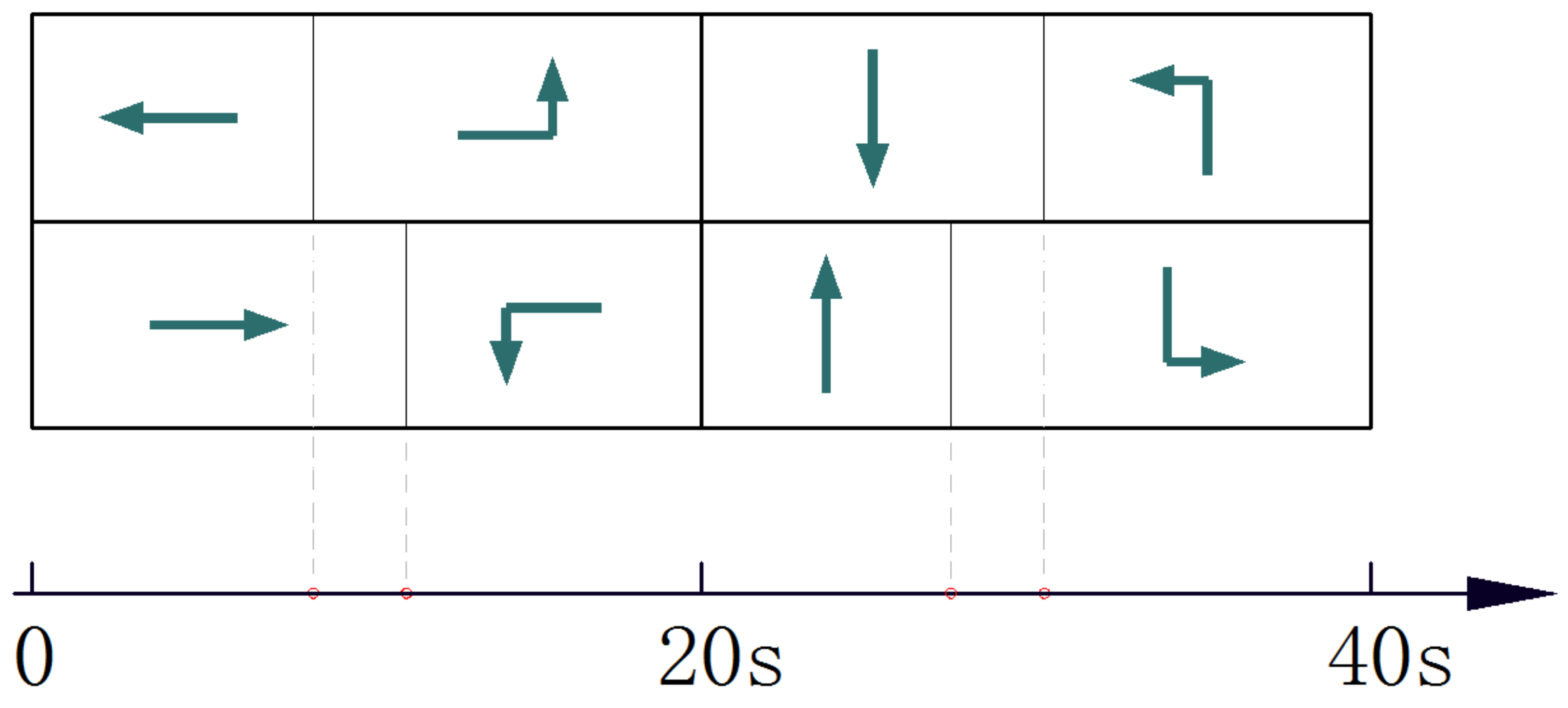}}	
	\caption{An example dual-ring signal cycle.}
	\label{F:dual}%
\end{figure}

We now perform experiments on the four-intersection arterial depicted in Fig.~\ref{F:Arterial}. We assume the same (mean) arrival rates to the four intersections and employ a max-weight control scheme.  Although the control scheme is decentralized, we allow communication between the vehicles and all four intersections and between the intersection controllers.  We assume the cycle lengths are the same for the four intersections (40s), the yellow indication is 1.8s, and the off-set time between adjacent intersections is 20s (we coordinate the green times for good signal progression).  An example dual-ring cycle is shown in Fig.~\ref{F:dual}.  The vehicle flux function $\phi_{a,b}(t,\pi^j)$ used is based on Newell's non-linear flow-density relation \cite{newell1961nonlinear}
\begin{equation}
	\phi = v_{\rm{f}}\rho  - v_{\rm{f}}\rho \exp\left(\frac{w}{v_{\rm{f}}}\Big( 1-\frac{\rho_{\rm{jam}}}{\rho} \Big)\right),
	\label{E:v-rho}
\end{equation}
where $\rho$ is the traffic density, $v_{\rm{f}} = 60~\rm{km/h}$ is the free-flow speed, $w = 25~\rm{km/h}$ is the backward wave speed, and $\rho_{\rm{jam}} = 133~\rm{veh/\big( km \cdot lane\big)}$ is the jammed traffic density.

Using the same random seeds, we test three scenarios. The first scenario (Coor-Pay) employs our proposed direct-transaction mechanism with signal coordination. The second scenario (Iso-Pay) also employs a direct-transaction mechanism but without signal coordination (i.e., the intersections are isolated). The third scenario (Coor-TV) assumes no economic instruments but the signals are coordinated.

\begin{figure}[h!] 
	\centering
	\resizebox{0.32\textwidth}{!}{%
		\includegraphics{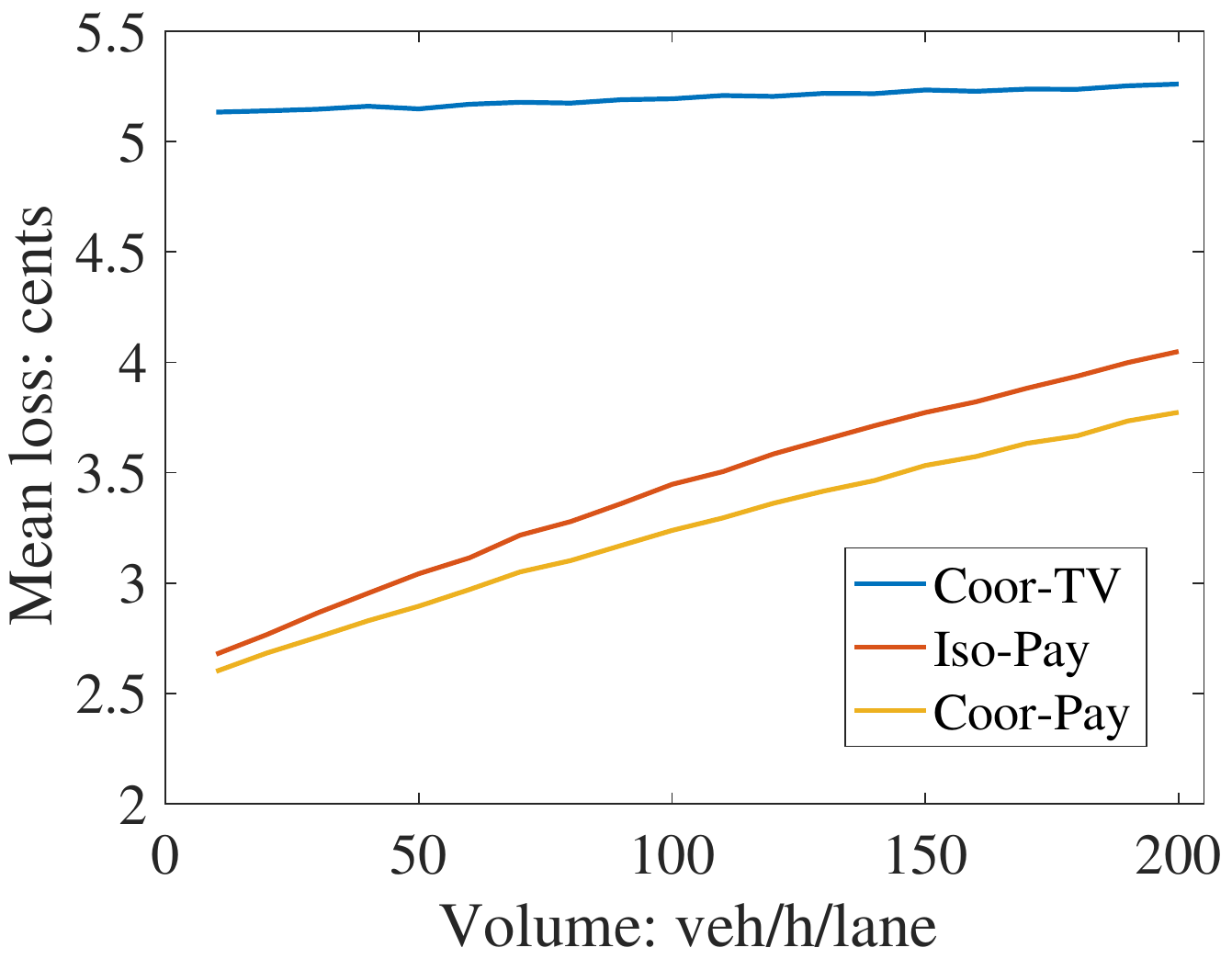}}	
	\caption{Mean loss of all arterial vehicles: bottom (orange) curve: Coor-Pay, middle (red) curve: Iso-Pay, top (blue) curve: signal coordination without economic instruments (baseline: free-flow passing).}
	\label{F:Artery_3curves}%
\end{figure}

\begin{figure}[h!] 
	\centering
	\resizebox{0.4\textwidth}{!}{%
		\includegraphics{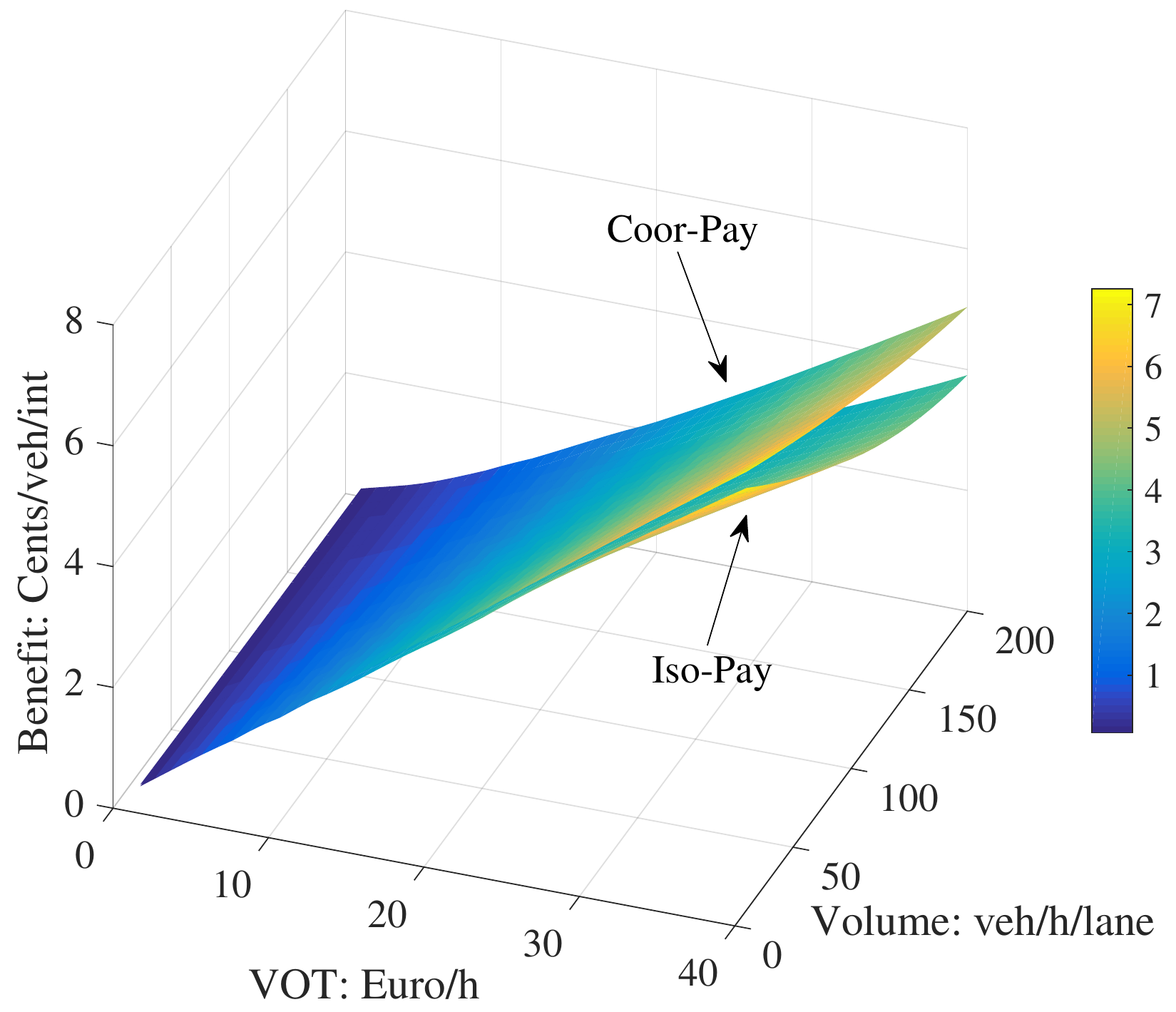}}	
	\caption{Expected benefits of Coor-Pay and Iso-Pay.}
	\label{F:Artery_3D}%
\end{figure}

We calculate the mean loss \eqref{E:lv} of all vehicles at the four intersections under different volumes  The results are shown in Fig.~\ref{F:Artery_3curves}. Clearly, Coor-Pay out-performs the other two under all tested volumes, and the third scenario, which only includes signal coordination performs far worse than the other two over all volumes.  We further compare mean benefits \eqref{E:bv} for the first two (Coor-Pay and Iso-Pay) over varying vehicle VOTs in Fig.~\ref{F:Artery_3D} and see that the differences are larger for vehicles with higher VOTs, but the difference is generally not very large.

\section{Conclusion}
This paper designed a simple and efficient cooperative framework for intersection control, which is compatible with and complements a variety of existing real-time (adaptive) control methods.  The approach employs a cooperative games framework, namely, transfer of utility, to calculate the payments.  In contrast to most popular intersection control schemes with economic instruments (e.g., auctions), in the proposed setting, winners compensate the losers directly.  We also demonstrated how the proposed payment mechanism can be combined with a variety of existing intersection control schemes.  We tested the proposed scheme using simulation experiments for an isolated intersection employing a reservation-based scheme and a signalized arterial employing max-weight control.  In the former case, it was shown that the proposed economic instrument outperforms second best auctions, and in the latter case, we demonstrated the efficacy of economic instruments compared to coordinated signal control without economic instruments.

\section*{Acknowledgment}
This work was supported by the NYUAD Center for Interacting Urban Networks (CITIES), funded by Tamkeen under the NYUAD Research Institute Award CG001 and by the Swiss Re Institute under the Quantum Cities\textsuperscript{TM} initiative. The views expressed in this article are those of the authors and do not reflect the opinions of CITIES or its funding agencies.

\appendix
\gdef\thesection{Appendix \Alph{section}}

	
	
	
\bibliographystyle{plainnat}
\bibliography{literature}
	
	
	
	
	
	

\end{document}